\documentclass[twocolumn]{aastex631}
\usepackage{amsmath,amssymb}
\usepackage{graphicx}
\graphicspath{{./FIGURES2/}}
\usepackage{hyperref}
\hypersetup{colorlinks=true,citecolor=blue,linkcolor=blue,urlcolor=blue}
\usepackage{bm}
\usepackage{here}
\usepackage{natbib}

\begin{document}

\title{The Effect of the Chromospheric Temperature on Coronal Heating}

\author{Haruka Washinoue}
\affiliation{School of Arts and Sciences, The University of Tokyo, 3-8-1, Komaba, Meguro, Tokyo 153-8902, Japan}

\author{Munehito Shoda}
\affiliation{National Astronomical Observatory of Japan, National Institutes of Natural Sciences, 2-21-1 Osawa, Mitaka, Tokyo, 181-8588, Japan}
\affiliation{Department of Earth and Planetary Science, The University of Tokyo, 7-3-1, Hongo, Bunkyo, Tokyo, 113-0033, Japan}

\author{Takeru K. Suzuki}
\affiliation{School of Arts and Sciences, The University of Tokyo, 3-8-1, Komaba, Meguro, Tokyo 153-8902, Japan}
\affiliation{Department of Astronomy, The University of Tokyo, 7-3-1, Hongo, Bunkyo, Tokyo, 113-0033, Japan}

\begin{abstract}
Recent observational and numerical studies show a variety of thermal structures in the solar chromosphere.
Given that the thermal interplay across the transition region is a key to coronal heating, it is worth investigating how different thermal structures of the chromosphere yield different coronal properties.
In this work, by MHD simulations of Alfvén-wave heating of coronal loops, we study how the coronal properties are affected by the chromospheric temperature.  To this end, instead of solving the radiative transfer equation, we employ a simple radiative loss function so that the chromospheric temperature is easily tuned. 
When the chromosphere is hotter, because the chromosphere extends to a larger height, the coronal part of the magnetic loop becomes shorter, which enhances the conductive cooling.
A larger loop length is therefore required to maintain the high-temperature corona against the thermal conduction.
From our numerical simulations we derive a condition for the coronal formation with respect to the half loop length $l_{\rm loop}$ in a simple form: $l_{\rm loop} > a T_{\rm min} + l_{\rm th}$,  where $T_{\rm min}$ is the minimum temperature in the atmosphere, and parameters $a$ and $l_{\rm th}$ have negative dependencies on the coronal field strength.
Our conclusion is that the chromospheric temperature has a nonnegligible impact on coronal heating for loops with small lengths and weak coronal fields.
In particular, the enhanced chromospheric heating could prevent the formation of the corona.
\end{abstract}


\section{Introduction}

The solar corona is the outer part of the atmosphere that consists of hot plasma in excess of one million Kelvin. 
Its dynamic activity has been often captured by the observation of soft X-rays and extreme ultraviolet (EUV) that are dominantly emitted from closed magnetic structures \citep{Reale2014}.
In other words, coronal heating regulates the stellar X-ray and EUV emission that significantly affects the evolution of protoplanetary disks and planetary atmospheres \citep{Vidal-Madjar2003, Sanz-Forcada2011, Ehrenreich2015, Nakatani2018}. 
Therefore, solving the coronal heating problem is one of the most important problems in astronomy.

One important factor in solving the coronal heating problem is the energy transfer from the photosphere to the corona. 
Among various forms of energy transfer, Alfv\'{e}n waves are expected to be dominant because they experience neither shock formation nor refraction \citep{Alfven1947, Osterbrock1961}.
The kinetic energy of the surface convection is carried in the form of Poynting flux of the Alfv\'{e}n waves.
It is finally converted into the thermal energy via the wave dissipation to heat the hot corona.
Alfv\'{e}n waves dissipate through multiple processes such as nonlinear mode conversion to compressible waves \citep{Hollweg1982a, Kudoh1999, Suzuki2005, Vasheghani2012}, turbulent cascading \citep{Goldreich1995, Matthaeus1999, Cranmer2007, Verdini2010, vanBallegooijen2016, Shoda2019}, phase mixing \citep{Heyvaerts1983, DeMoortel2000, Goossens2012}  and resonant absorption \citep{Erdelyi1995, Terradas2010, Goossens2011}.
Whatever the dissipation mechanisms are, a large amount of Alfv\'{e}nic Poynting flux needs to be transported to the corona.

Given that the bulk of the Poynting flux is likely to be transferred in the form of Alfv\'{e}n waves through the chromosphere, it is important to numerically investigate the nature of Alfv\'{e}n-wave propagation in the solar atmosphere.
In particular, a self-consistent calculation that covers from the photosphere to the corona provides us with a systematic understanding of the coronal heating from wave generation, energy transfer, and conversion to heat. 
Such simulations have been extensively developed for the heating of coronal loops \citep{Moriyasu2004, Antolin2010, Matsumoto2016, Matsumoto2018, Washinoue2019, Shoda2021} and the solar wind acceleration \citep{Suzuki2005, Suzuki2006, Matsumoto2012, Matsumoto2014, Shoda2018a, Sakaue2020, Matsumoto2021}. 
These theoretical works show that Alfv\'{e}n waves have a potential to feed a sufficient amount of energy into the corona.
 
Meanwhile, there has been little focus on the relation between the chromospheric structure and the coronal heating.
The dynamical, thermal, and magnetic properties of the chromosphere affect the Alfv\'{e}n-wave propagation in the chromosphere, and the coronal heating as well.
Recent observations show a highly dynamic and complicated nature of the chromosphere \citep{DePontieu2014, Tian2014, Yokoyama2018}. It is worth investigating the effect of different chromospheric conditions on the coronal heating.  

A key uncertain factor is the minimum temperature of the atmosphere, $T_{\rm min}$, which is located at the bottom of the chromosphere. 
$T_{\rm min} \approx 4200$ K is widely used as derived from the standard model of the static solar atmosphere \citep{Vernazza1981}.  
However, some observations indicate a lower value of $T_{\rm min}$ ($<4000$ K) based on the existence of carbon monoxide molecules in quiet-sun regions \citep{Noyes1972, Solanki1994, Ayres1996}. 
Recently, even cooler regions with $T\sim 3000$ K are inferred from the inversion of the chromospheric spectral lines retrieved by the Interface Region Imaging Spectrograph (IRIS) and the Atacama Large Millimeter/submillimeter Array (ALMA; \cite{daSilva2020}).
Cool chromospheric materials are also found in numerical simulations. For example, a radiation MHD simulation with nonlocal thermodynamic equilibrium effects by \cite{Leenaarts2011} shows the formation of cool gas with $T_{\rm min}< 2000$ K in the solar chromosphere.  
We note, on the other hand, that the chromospheric temperature could be hotter than what the classical model predicts under the presence of ambipolar diffusion \citep{Khomenko2012, MartinezSykora2015, Martinez-Sykora2017}. Therefore, in spite of recent progresses in the chromospheric modeling, the chromospheric temperature still remains uncertain.

Given the uncertainty of the chromospheric temperature, we study the effect of $T_{\rm min}$ on the coronal heating. 
For this purpose, we carry out one-dimensional MHD simulations for coronal loops and compare the loop profiles and radiative luminosities over a wide range of $T_{\rm min}$.

This paper is organized as follows.
In Section \ref{sec2}, we describe the numerical model and settings of the chromospheric temperature for our simulations.
The results for the dependence of the coronal properties on $T_{\rm min}$ are shown in Section \ref{sec3}. 
We present further analysis and discussions in Section \ref{sec4} and summarize the paper in Section \ref{sec5}. 

\section{Simulation}
\label{sec2}

\subsection{Basic Setup}
\label{sec21}

The stellar parameters in our simulations are adopted for those of the present Sun with mass $M_{\odot}$ and radius $R_{\odot}$. 
The effective temperature is $T_{\rm eff}=5780$ K, the photospheric density is $\rho_{\rm ph}=2.5\times 10^{-7}$ g cm$^{-3}$ and the magnetic field strength at the photosphere is fixed to $B_{\rm ph}=1.58$ kG, which gives the equipartition between the magnetic energy and the thermal energy \citep{Cranmer2011, Suzuki2018, Washinoue2019}.

A loop is modeled by a one-dimensional semicircular magnetic flux tube anchored in the photosphere. We take coordinates $(s, \perp_1, \perp_2)$, where $s$ is along the loop and the other two components are perpendicular to $s$. 
We adopt the 1.5-dimensional approximation: 
\begin{align}
\frac{\partial}{\partial \perp_1}=\frac{\partial}{\partial \perp_2}=0.
\label{eq1}
\end{align}

\subsection{Flux Tubes}
\label{sec22}

\begin{figure}[t]
\begin{centering}
\includegraphics[width=85mm]{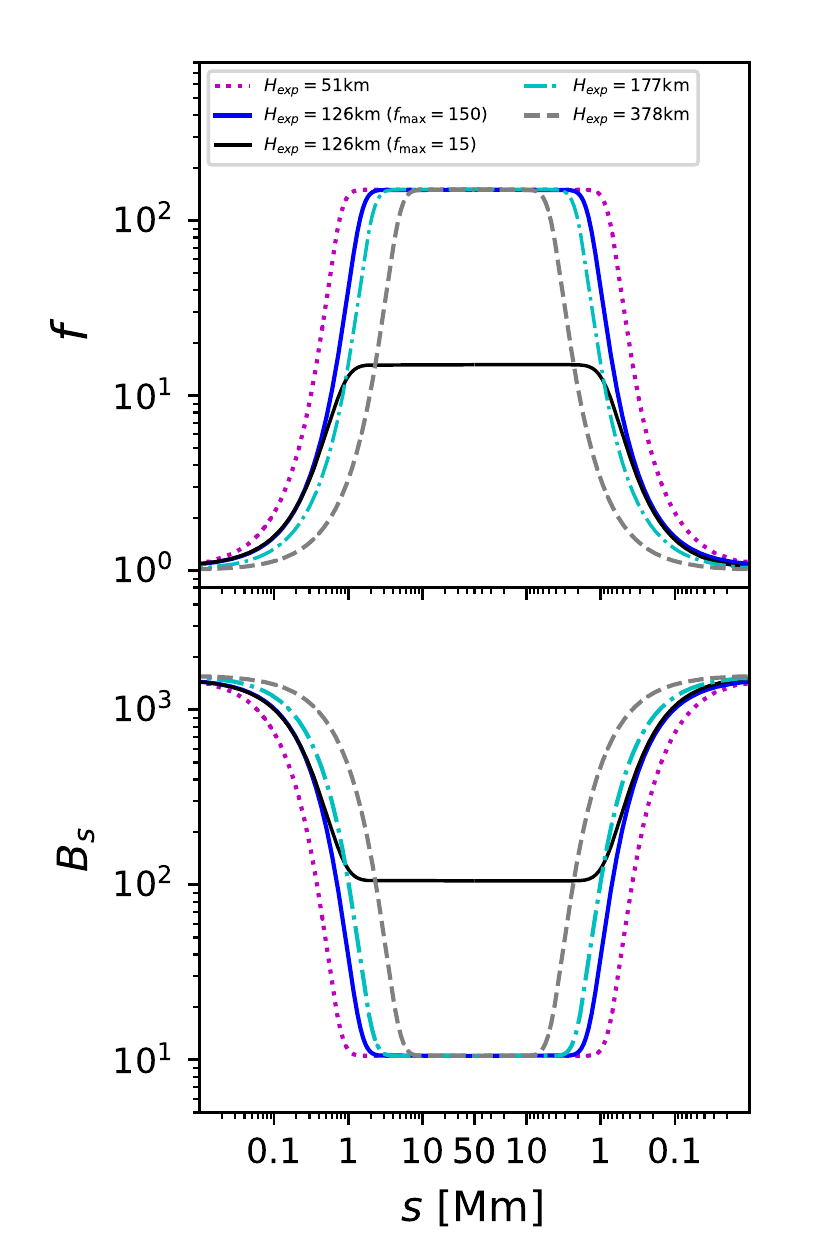}
\end{centering}
  \caption{Profiles of the expansion factor $f(s)$ (top) and $B_s$ (bottom) for different $H_{\rm exp}$ and $f_{\rm max}$. 
  The black solid line corresponds to $f_{\rm max}=15$ and the others to $f_{\rm max}=150$.}
  \label{fig1}
\end{figure}

Because the flux tube expands with the height in the chromosphere \citep{Cranmer2005, Ishikawa2021}, we formulate the expansion factor as
\begin{align}
f(s) = \frac{f_{\mathrm{max}}}{2} \left\{\mathrm{tanh} \left[ a \left(\frac{-|l_{\rm loop}-s|/R_{\odot}+b}{2l_{\rm loop}/\pi R_{\odot}}+\frac{\pi}{4}\right)\right] + 1 \right\},
\label{eq2}
\end{align}
where $f_{\rm max}$ is the value of $f(s)$ at the loop top.  $f_{\rm max}=150$ is adopted for our standard cases, which gives the coronal magnetic field strength $B_{\rm cor}=B_{\rm ph}/f_{\rm max}=10.5$ G. 
This value corresponds to the average coronal field strength \citep{Long2013,Reale2014}, and then,  the standard cases target magnetic loops in quiet-Sun regions.
We also run simulations with the strong coronal field of $B_{\rm cor}=105$ G aiming at loops in active regions by reducing $f_{\rm max}$ to 15 in Section \ref{sec43}. 
$l_{\rm loop}$ is the half loop length. 
Parameters $a$ and $b$ define the geometry of the flux tube.
These are formalized, as a function of the height of the expansion $H_{\rm exp}$, as follows:
\begin{align}
a = -\frac{l_{\rm loop}}{5\pi H_{\rm exp}} \mathrm{arctanh}\left(\frac{2}{f_{\rm max}}-1\right),
\label{eq3}
\end{align}
\begin{align}
b =  \frac{1}{R_{\odot}} \left( \frac{l_{\rm loop}}{2} - 10 H_{\rm exp} \right).
\label{eq4}
\end{align}
In particular, in the fiducial setting, $H_{\rm exp}$ is set to be the isothermal pressure scale height at $T = T_{\rm off}$, where $T_{\rm off}$ is the cutoff temperature of the radiative cooling (see Section \ref{sec24}).
The profiles of $f(s)$ and $B_s = B_{\rm ph}/f(s)$ ($s$ component of the magnetic field) are displayed in Figure \ref{fig1}. 
The blue solid line ($H_{\rm exp}=126$ km and $f_{\rm max}=150$) represents the fiducial model for $l_{\rm loop}=50$ Mm and $T_{\rm off}=5000$ K. 
$f(s)$ reaches $f_{\rm max}$ at 1.8 Mm in this case.
Other $H_{\rm exp}$ cases are used to discuss the effect of the geometry of the flux tube in Section \ref{sec42}.

\subsection{Equations}
\label{sec23}

We numerically solve time-dependent MHD equations with phenomenological  turbulent dissipation \citep{Shoda2018a}:
\begin{align}
 \frac{\partial \rho}{\partial t} + \frac{1}{f} \frac{\partial}{\partial s} (\rho v_s f) = 0,
 \label{eq5}
\end{align}
\begin{align}
\begin{split}
 \frac{\partial}{\partial t} (\rho v_s) &= - \frac{1}{f} \frac{\partial}{\partial s}\left[ \left( \rho v_s^2 + P + \frac{B_\perp^2}{8\pi}\right) f \right] \\
 &+ \frac{1}{f} \left(P + \frac{\rho v_\perp ^2}{2}\right) \frac{\partial f}{\partial s}- \frac{\rho GM_{\odot}}{R^2}\mathrm{cos}\left(\frac{\pi s}{2l_{\mathrm{loop}}}\right), 
  \end{split}
  \label{eq6}
\end{align}
\begin{align}
\begin{split}
\frac{\partial}{\partial t} (\rho \bm{v}_\perp) = &- \frac{1}{f^{3/2}}\frac{\partial}{\partial s}\left[ \left(\rho v_s \bm{v_\perp} -  \frac{B_s \bm{B_\perp}}{4\pi}\right)f^{3/2}\right]  \\
 &-\hat{\bm{\eta}}_1 \cdot \rho \bm{v_\perp} - \hat{\bm{\eta}}_2 \cdot \sqrt{\frac{\rho}{4\pi}} \bm{B_{\perp}}, 
  \end{split}
   \label{eq7}
\end{align}
\begin{align}
\begin{split}
 \frac{\partial \bm{B}_\perp}{\partial t} = &\frac{1}{\sqrt{f}} \frac{\partial}{\partial s}\left[(\bm{v_\perp} B_s - v_s \bm{B_{\perp}}) \sqrt{f}\right]\\ 
 & - \hat{\bm{\eta}}_1 \cdot \bm{B_{\perp}} - \hat{\bm{\eta}}_2
  \cdot \sqrt{4\pi \rho} \bm{v}_\perp,
   \end{split}
   \label{eq8}
\end{align}

\begin{align}
\begin{split}
 &\frac{\partial}{\partial t}\left(\rho e+\frac{\rho v^2}{2}+\frac{B_{\perp}^2}{8\pi} \right)  \\
 &+ \frac{1}{f}\frac{\partial}{\partial s} \left[\left(\rho e+\frac{\rho v^2}{2}+P+\frac{B_{\perp} ^2}{4\pi}\right)v_s f -\frac{B_s}{4\pi}\left(\bm{B_\perp}\cdot\bm{v_\perp}\right) f \right] \\
 & + \frac{\rho GM_{\odot}}{R^2}v_s \mathrm{cos}\left(\frac{\pi s}{2l_{\mathrm{loop}}}\right) + \frac{1}{f}\frac{\partial}{\partial s} (F_c f) + q_R = 0
\end{split}
\label{eq9}
\end{align}
and
\begin{align}
P = \frac{\rho k_B T}{\mu m_H},
\label{eq10}
\end{align}
where $G$ is the gravitational constant. 
$R$ is the height measured from the center of a star, which is related to $s$ as $R=R_{\odot}+(2l_{\rm loop}/\pi)\mathrm{sin}(\pi s/2l_{\rm loop})$.
$e=P/(\gamma -1)\rho$ is the internal energy per unit mass, where $\gamma = 5/3$ is the ratio of specific heats.
$k_B$ is the Boltzmann constant, and $m_H$ is the atomic mass unit.
$\mu$ is the mean molecular weight; we adopt $\mu=1.2$ for $T\le 6\times 10^3$ K and $\mu=0.6$ for $T\ge 10^4$ K, where we smoothly interpolate $\mu$ in the intermediate range, $6\times 10^3$ K $< T < 10^4$ K. 

$\hat{\bm{\eta}}_1$ and $\hat{\bm{\eta}}_2$ in Equations (\ref{eq7}) and (\ref{eq8}) are diagonal matrices for the diffusion terms described as follows \citep{Shoda2018a};
\begin{align}
\hat{\bm{\eta}}_1 = 
\begin{pmatrix}
\dfrac{c_d}{4\lambda} \left( \left| z_{\perp_1} ^+\right| + \left| z_{\perp_1} ^-\right|\right) & 0 \\
0 & \dfrac{c_d}{4\lambda} \left( \left| z_{\perp_2} ^+\right| + \left| z_{\perp_2} ^-\right|\right) \\
\end{pmatrix},
\label{eq11}
\end{align}
\begin{align}
\hat{\bm{\eta}}_2 = 
\begin{pmatrix}
\dfrac{c_d}{4\lambda} \left( \left| z_{\perp_1} ^+\right| - \left| z_{\perp_1} ^-\right|\right) & 0 \\
0 & \dfrac{c_d}{4\lambda} \left( \left| z_{\perp_2} ^+\right| - \left| z_{\perp_2} ^-\right|\right) \\
\end{pmatrix},
\label{eq12}
\end{align}
where $\lambda$ is the perpendicular correlation length of the Alfv\'{e}nic turbulence. $z_{\perp} ^{\pm}=v_{\perp} \mp B_{\perp}/\sqrt{4\pi \rho}$ are the Els$\ddot{\rm a}$sser variables \citep{Elsasser1950}.
We assume that $\lambda$ increases with the width of the flux tube:
\begin{align}
\lambda = \lambda_0 \sqrt{f(s)},
\label{eq13}
\end{align}
where $\lambda_0$ represents the correlation length at the photosphere, and we adopt $\lambda_0 = 100$ km  as a typical size of intergranular lanes. 
$c_d$, the dimensionless parameter to characterize the turbulent dissipation, is set to $c_d=0.1$, following \cite{VanBallegooijen2017} and \cite{Verdini2019}.

In Equation (\ref{eq9}), $F_c$ is the thermal conductive flux that is described as
\begin{align}
F_c= -\kappa T^{5/2} \frac{\partial T}{\partial s},
\label{eq14}
\end{align}
where $\kappa = 10^{-6}$ erg cm$^{-1}$ s$^{-1}$ K$^{-7/2}$ is the Spitzer conductivity \citep{Spitzer1953}.
$q_R$ denotes the radiative cooling rate per unit volume.
In the hot ($T \geq 10^4$ K) and tenuous ($\rho \leq 10^{-16}$ g  cm$^{-3}$) region, we employ the optically thin cooling for $q_R$:
\begin{align}
q_R=\Lambda (T) n n_e,  
\label{eq15}
\end{align}
where $\Lambda (T)$ is the cooling function, and we use the cooling table from \cite{Sutherland1993}. $n$ and $n_e$ are ion and electron number densities, respectively. 
In the other regions, we adopt a simple prescription based on the empirical cooling rate for the solar chromosphere \citep{Anderson1989}, 
\begin{align}
q_R=4.5\times 10^9 \rho,
\label{eq16}
\end{align}
where $q_R$ and $\rho$ are measured in cgs units. 
Because we artificially tune the chromospheric temperature by quenching the radiative cooling (Section \ref{sec24}), solving the radiative transfer equation would be an excessive effort.

\subsection{Chromospheric Temperature}
\label{sec24}

We artificially set up different chromospheric temperatures to investigate the impact of the chromospheric structure on the corona. 
To vary the temperature in the chromosphere, we introduce a cutoff temperature for the radiative cooling, $T_{\rm off}$; when $T<T_{\rm off}$, $q_R$ is turned off. 
$T_{\rm off}$ determines the minimum temperature, $T_{\rm min}$, and the temperature structure in the chromosphere.
Accordingly, it also governs the density profile through the stratification of the atmosphere, which affects the propagation and reflection of Alfv\'{e}n waves. 

\subsection{Initial Condition and Wave Injection}
\label{sec25}

We initially set a static ($\bm v=0$) and isothermal ($T=T_{\rm eff}$) atmosphere all along the loop.
Velocity perturbations are injected from the photosphere in all the three directions to excite shear Alfv\'{e}n waves and acoustic waves.
We adopt the same rms amplitude of $\delta v = 1.0$ km s$^{-1}$ for each component, which is comparable to the observed photospheric motions \citep{Matsumoto2010}.
The spectral shape is proportional to $\omega^{-1}$ in a range between $\omega_{\rm min}=2.0\times 10^{-3}$ Hz and $\omega_{\rm max} =3.3\times 10^{-2}$ Hz, where $\omega$ is the frequency of the photospheric fluctuation.

\subsection{Numerical Resolution}
\label{sec26}

The grid size is gradually increased from the photosphere to the corona in accordance with the variation in the Alfv\'{e}n speed to resolve Alfv\'{e}n waves at different locations.
We set the grid size as
\begin{align}
\Delta s = 
 \begin{cases}
 {\Delta s_{\rm min} \hspace{2mm}(s\leq s_l)} \\
 {\frac{\Delta s_{\rm max}}{2}\left\{\mathrm{tanh} \left[\alpha s\left(\frac{s_l - s}{s_l}+1\right)\right]+1\right\}} \\
 {(s> s_l),}
  \end{cases}
\end{align}
where $\Delta s_{\rm min}=5$ km, $\Delta s_{\rm max}=80$ km, $s_l=3000$ km, and $\alpha=4.5\times 10^{-4}$.
We also perform higher-resolution simulations with $\Delta s_{\rm min}=2$ km for some cases to examine the dependence on the numerical resolution (Section \ref{sec48}).
In this study, the second-order Godunov method is used to calculate the compressible waves, and the method of characteristics is used for incompressible waves \citep{Stone1992,Suzuki2005}.

\begin{figure}[!t]
\begin{centering}
\includegraphics{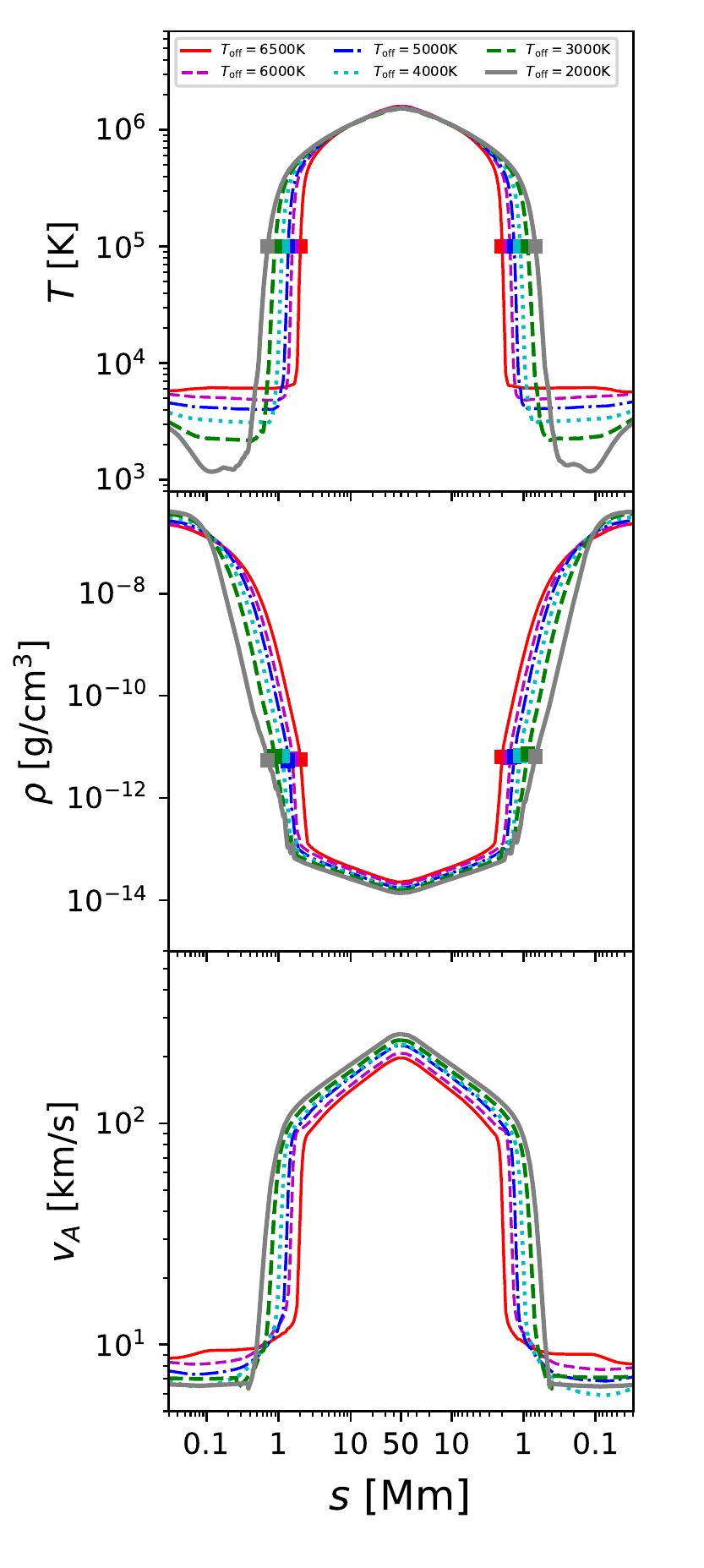}
\end{centering}
  \caption{Time-averaged loop profiles of the temperature (top), density (middle), and Alfv\'{e}n speed (bottom) for $l_{\rm loop}=50$ Mm with different $T_{\rm off}$. The squares mark the positions at $T=10^5$ K.}
  \label{fig2}
\end{figure} 

\begin{figure}[t]
\begin{centering}
\includegraphics[width=80mm]{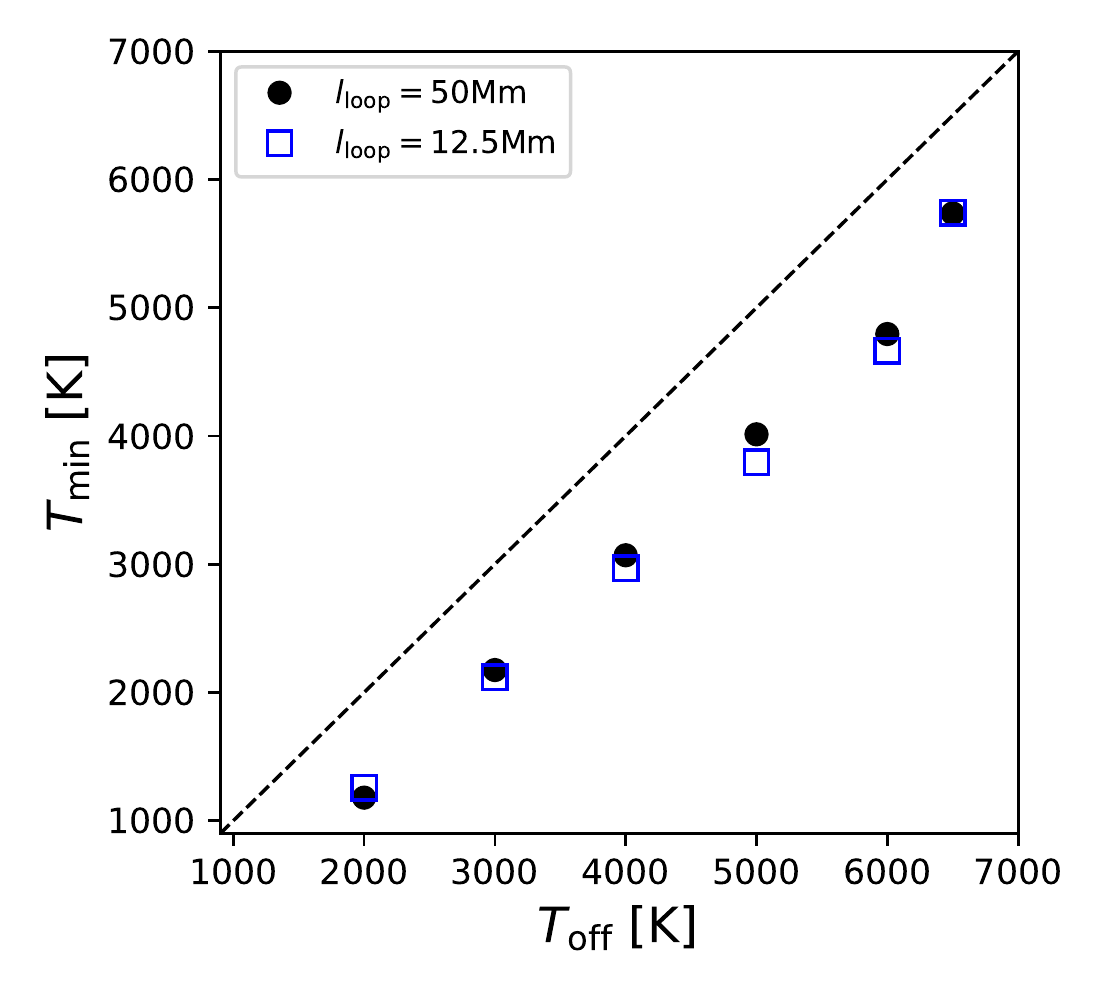}
\end{centering}
  \caption{$T_{\rm min}$ vs. $T_{\rm off}$ in the cases of $l_{\rm loop}=50$ Mm and $12.5$ Mm. The dashed line denotes $T_{\rm min}=T_{\rm off}$.}
  \label{fig3}
\end{figure} 

\section{Results}
\label{sec3}

We perform simulations for six different cutoff temperatures in the range of 2000 K $\le T_{\rm off} \le$ 6500 K to examine the effect of the chromospheric structure on the corona. 
We also test cases with different half loop lengths, $l_{\rm loop}= 10, 12.5, 15, 20, 25, 50,$ and $100$ Mm. 

In our simulation, velocity perturbations are injected with the broadband spectrum (Section \ref{sec25}).
Alfv\'{e}n waves, which are driven by the high-frequency transverse fluctuations, propagate upward and play a role in alternative current (AC) heating; they dissipate via turbulent cascade \citep{Shoda2021} and nonlinear mode conversion to compressible waves \citep{Moriyasu2004} to heat the gas.
The low-frequency fluctuations, whose periods are longer than the Alfv\'{e}n crossing time, are responsible for direct current (DC) heating through the diffusion terms in Equations (\ref{eq7}) and (\ref{eq8}).

All the simulated coronal loops reached quasi-steady states within approximately $1.5$ hours after the start of the simulations. We take the time average from $t=3$ hr to $t=15$ hr to compare the general properties of different cases.

\begin{figure}[htb]
\begin{centering}
\includegraphics{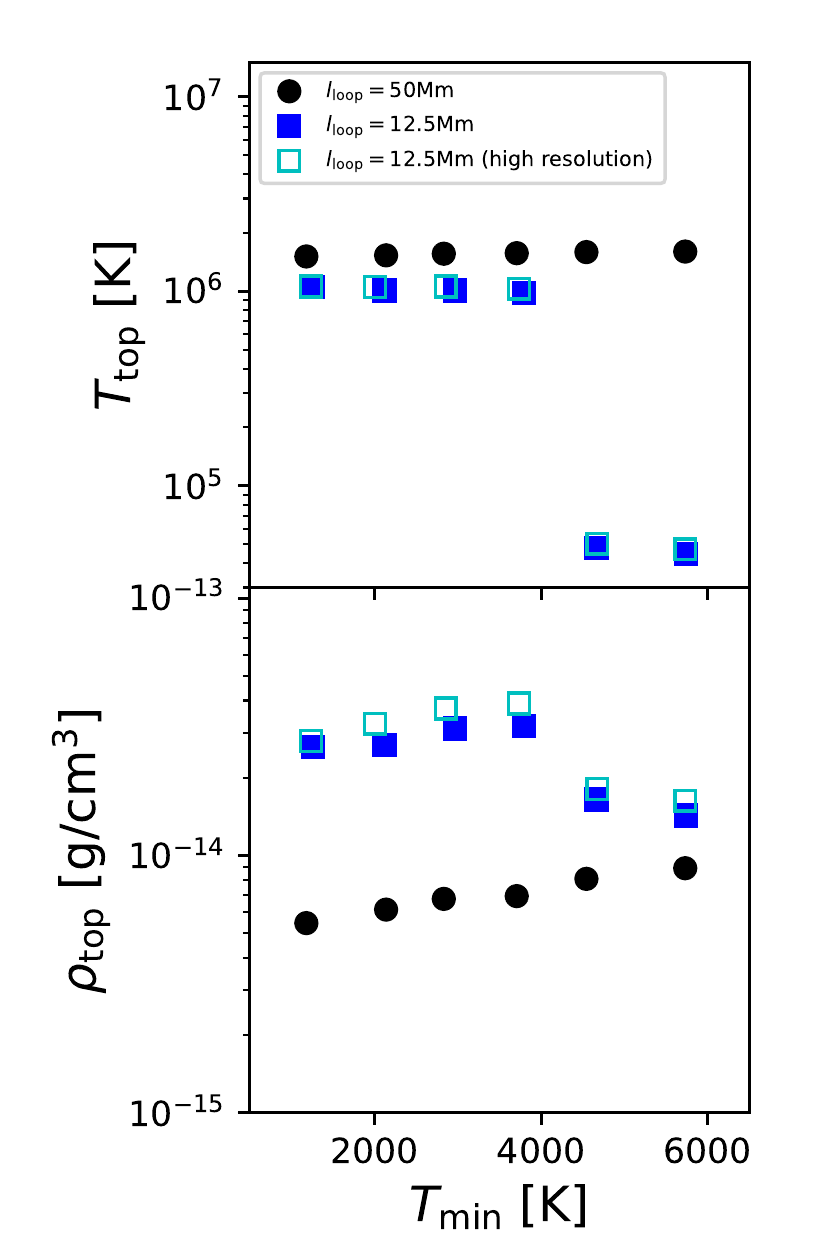}
\end{centering}
  \caption{Temperature (top) and density (bottom) at the loop top with $T_{\rm min}$ for $l_{\rm loop}=50$ Mm (black circles) and $l_{\rm loop}=12.5$ Mm (blue squares). Cyan squares are the cases with 2 km for resolution at the transition region for $l_{\rm loop}=12.5$ Mm.}
  \label{fig4}
\end{figure}
\begin{figure}[h]
\includegraphics{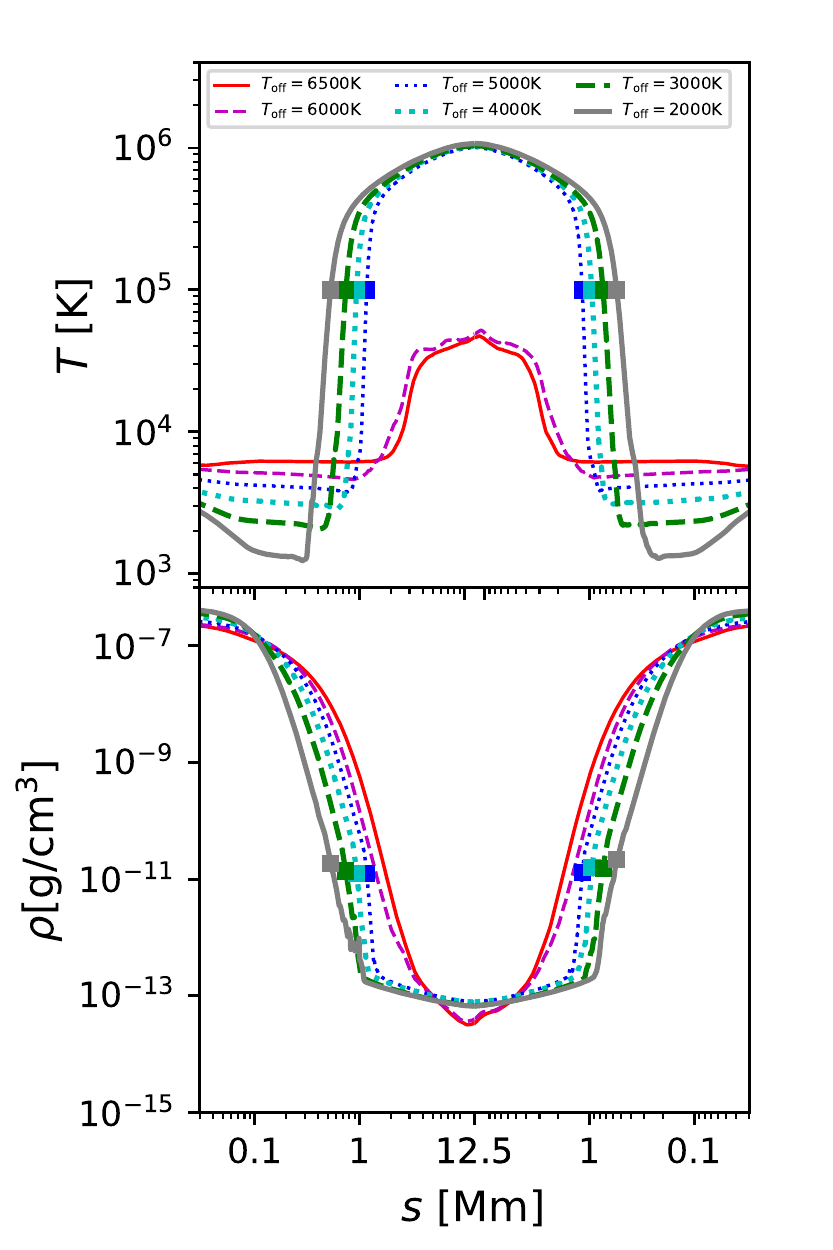}
  \caption{Time-averaged loop profiles of temperature (top) and density (bottom) for $l_{\rm loop}=12.5$ Mm with different $T_{\rm off}$. The squares mark the positions at $T=10^5$ K.}
  \label{fig5}
\end{figure} 

\subsection{Loop Profiles}
\label{sec31}
Figure \ref{fig2} shows the time-averaged profiles for the cases with $l_{\rm loop}=50$ Mm. 
It is clearly seen that different $T_{\rm off}$s give different chromospheric structures.    
Figure \ref{fig3} displays the relation between $T_{\rm off}$ and $T_{\rm min}$ for the cases with $l_{\rm loop} = 12.5$ and 50 Mm, where $T_{\rm min}$ is the time-averaged minimum temperature.
$T_{\rm min}$ is lower than $T_{\rm off}$ by $800-1400$ K because of the adiabatic cooling.
In high-$T_{\rm off}$ cases, the temperature decreases only weakly below the chromosphere (top panel of Figure \ref{fig2}) because of the strong quenching of radiative cooling.
As a result, the gas density slowly decreases with increasing altitude due to the extended scale height.

The transition region is governed by the energy balance between thermal conduction and radiative cooling.
As they have different dependencies on the density, for the fixed input energy and loop length, its location is primarily determined by the density, in spite of different $T_{\rm off}$.
In fact, the middle panel of Figure \ref{fig2} shows that the density of the transition region, which is represented by $\rho$ at $T=10^5$ K (squares), is almost independent from $T_{\rm min}$ (see also Section \ref{sec41}).
Therefore, the larger scale heights in the chromosphere give higher altitudes of the transition region.
For example, the height at $T=10^5$ K is $2000$ km for $T_{\rm off}=6500$ K, while it is $700$ km for $T_{\rm off}=2000$ K (squares in the top panel). 

On the other hand, the physical properties of the corona are similar for different $T_{\rm off}$ cases with $l_{\rm loop}=50$ Mm, although we see that the coronal temperature and density are slightly lower in lower-$T_{\rm off}$ cases. 
This can be clearly seen in the black circles of Figure \ref{fig4}, which shows the temperature and density at the loop top, $T_{\rm top}$ and $\rho_{\rm top}$ as a function of $T_{\rm min}$.
We note that the data in Figure \ref{fig4} are plotted against $T_{\rm min}$ (minimum temperature), which is lower than $T_{\rm off}$ (radiation cutoff temperature; Figure \ref{fig3}).
The weak dependence on $T_{\rm min}$ is a consequence of two conflicting effects. 
First, the rapid decline in the density in low-$T_{\rm off}$ cases results in the steep increase in the Alfv\'{e}n speed (middle and bottom panels of Figure \ref{fig2}), which enhances the reflection of Alfv\'{e}n waves  in the chromosphere \citep{An1990, Velli1993, Cranmer2005,Verdini2012,Suzuki2013}.
Therefore, a smaller fraction of the input Alfv\'{e}n waves can feed energy to the corona.
Second, as the chromosphere is relatively thin in low-$T_{\rm off}$ cases, the energy loss by radiative cooling at the low altitude is reduced. 
Consequently, these effects are canceled out, and a strong dependence on $T_{\rm min}$ does not appear.

The profiles of shorter loops ($l_{\rm loop}=12.5$ Mm) are shown in Figure \ref{fig5}.
The structures in the lower atmosphere are similar to those of the long-loop cases (Figure \ref{fig2}).
It shows, however, a different trend in terms of $T_{\rm min}$ in the upper atmosphere.
In particular, when $T_{\rm off} \geq 6000$ K, the million-Kelvin corona is not formed. 
In these cases, the chromosphere occupies a sizable fraction of the loop in length, which results in the decreased length in the coronal part. 
Therefore, there should be a condition for the coronal formation in terms of the coronal loop length, which is discussed in Section \ref{sec41}.
The dependences of $T_{\rm top}$ and $\rho_{\rm top}$ on $T_{\rm min}$ are also shown by the blue squares in Figure \ref{fig4}.  
$T_{\rm top}$ and $\rho_{\rm top}$ do not depend on $T_{\rm min}$ for $T_{\rm min} < 4000$ K, while they are much lower for $T_{\rm min}>4000$ K.
In short loops, the lower atmospheric structure significantly affects the physical condition in the upper atmosphere. We will discuss the details later in this section and Section \ref{sec41}.

\begin{figure}[htb]
\begin{centering}
\includegraphics{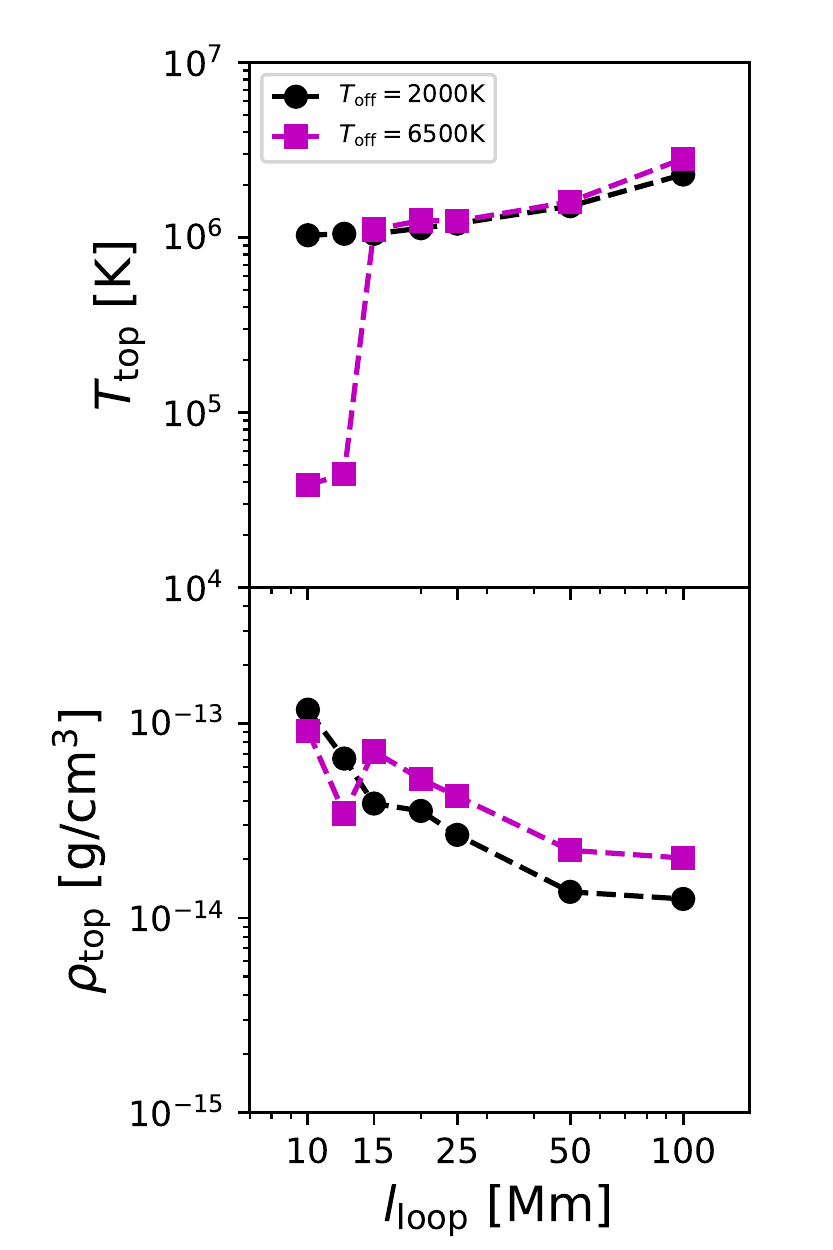}
\end{centering}
  \caption{Temperature (top) and density (bottom) at the loop top with loop length when $T_{\rm off}=2000$ K (black circles) and $6500$ K (pink squares).}
  \label{fig6}
\end{figure} 
\begin{figure}[htb]
\begin{centering}
\includegraphics{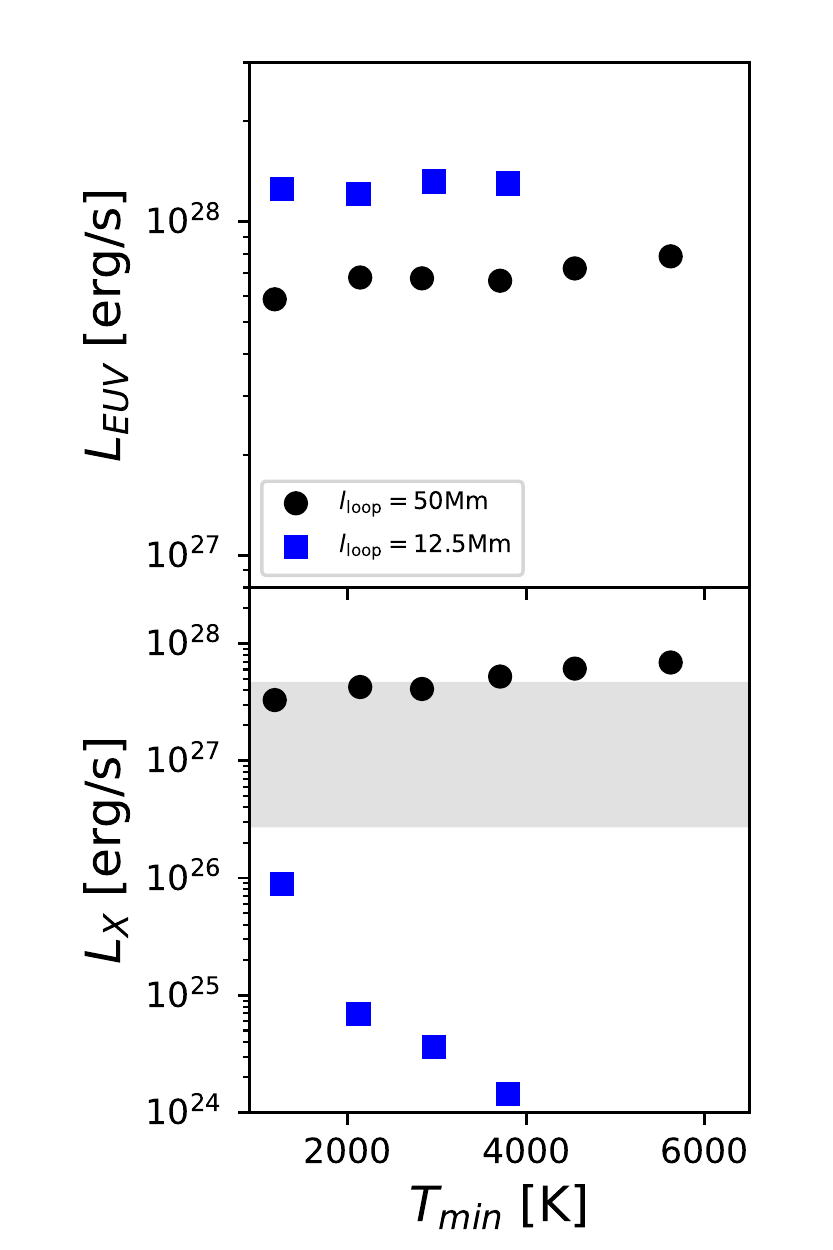}
\end{centering}
  \caption{EUV (top) and X-ray (bottom) luminosities with $T_{\rm min}$ for $l_{\rm loop}=50$ Mm and $12.5$ Mm. The shaded region represents the range of the solar X-ray luminosity obtained from the observation \citep{Peres2000, Johnstone2015}.}
  \label{fig7}
\end{figure} 

We also inspect the dependence on $l_{\rm loop}$ for different $T_{\rm off}$ cases. 
Figure \ref{fig6} shows $T_{\rm top}$ and $\rho_{\rm top}$ versus $l_{\rm loop}$ when $T_{\rm off}=6500$ K and $2000$ K.
It can be seen that longer loops generally give a higher temperature and lower density at the loop top, as predicted by the RTV scaling law \citep{Rosner1978}. 

$T_{\rm top}$ and $\rho_{\rm top}$ do not significantly depend on $T_{\rm min}$ in the cases with $l_{\rm loop} \geq 15$ Mm. 
In contrast, for the shorter loops with $l_{\mathrm{loop}}< 15$ Mm, the chromospheric temperature has a strong impact on the corona.  
The high-temperature corona is realized only in the low-$T_{\rm off}$ cases.
A thick chromosphere is formed in the high-$T_{\rm off}$ cases, so that the remaining length is no longer sufficient to maintain the high-temperature corona against thermal conduction. 
The gas at the loop top is heated only up to $(4-5)\times 10^5$ K when $T_{\rm off}=6500$ K in the loop of $l_{\rm loop}=12.5$ Mm, and it is even lower for $l_{\mathrm{loop}}=10$ Mm (see also Section \ref{sec41}).


\begin{figure}[t]
\begin{centering}
\includegraphics{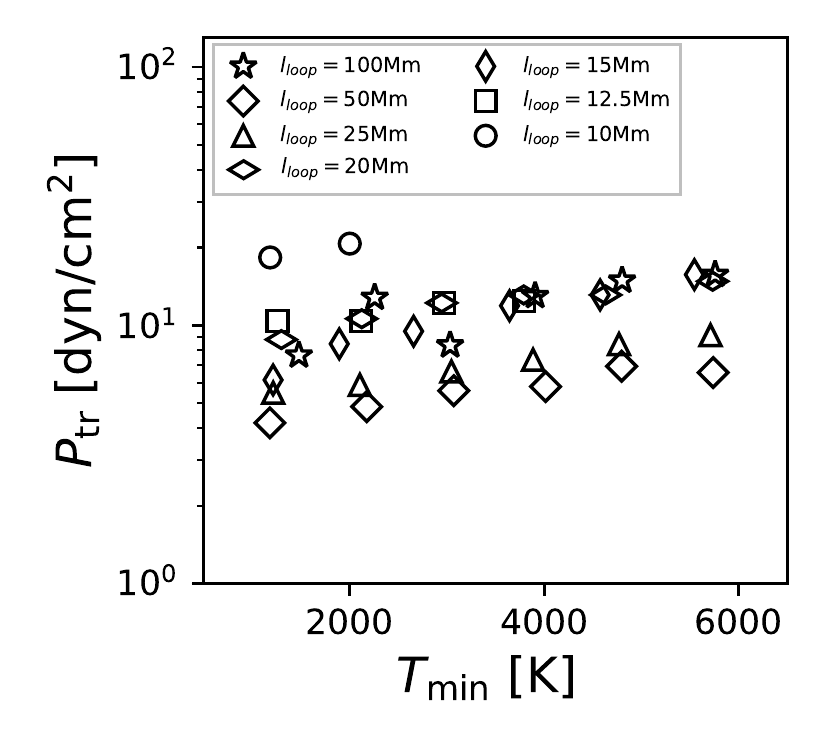}
\end{centering}
  \caption{$P_{\rm tr}$ with $T_{\rm min}$ for the cases that give $T_{\rm top}>0.5$ MK.  }
  \label{fig8}
\end{figure} 

\subsection{Radiative Luminosities}
\label{sec32}

 We estimate the radiative luminosity of a star whose surface is filled with the simulated loops \citep{Washinoue2019};
\begin{align}
&L = \frac{4 \pi R_{\odot}^2}{f_{\rm max} }  \int q_R f(s) ds .
\label{eq17}
\end{align}
We define $L_{\rm EUV}$ as the extreme-ultraviolet (EUV) luminosity from the gas in the temperature range of $1.5\times 10^5$ K$\leq T\leq1.1\times 10^6$ K and $L_{\rm X}$ as the (soft) X-ray luminosity with $T>1.1\times 10^6$ K, which respectively correspond to the energy ranges of 13.6 eV - 0.1 keV and $>0.1$ keV.

Figure \ref{fig7} shows the $T_{\rm min}-L_{\rm EUV}$ and $T_{\rm min}-L_X$ relations for $l_{\rm loop}=50$ Mm and $12.5$ Mm. 
The shaded region in the bottom panel represents the range of the solar X-ray luminosity obtained from the observation \citep{Peres2000, Johnstone2015}.
It can be seen that the dependence on $T_{\rm min}$ differs by the loop length.

When $l_{\rm loop}=50$ Mm, $L_{\rm EUV}$ is weakly affected by $T_{\rm min}$.
The spatial fraction of the EUV emission region is larger for smaller $T_{\rm min}$ because the transition region is located at a lower height.  
Meanwhile, the efficiency of emission is reduced because the density above the transition region is slightly lower (middle panel of Figure \ref{fig2}).
These two effects are canceled out to give a nearly constant $L_{\rm EUV}$ on $T_{\rm min}$.
On the other hand, $L_{\rm X}$ has a weak but nonzero dependence on $T_{\rm min}$; larger $T_{\rm min}$ values give larger $L_{\rm X}$. 
This is because the density with $T>1.1\times 10^6$ K near the loop top is higher for cases with higher $T_{\rm min}$ (Figure \ref{fig4}).  
\begin{figure}[t]
\begin{centering}
\includegraphics{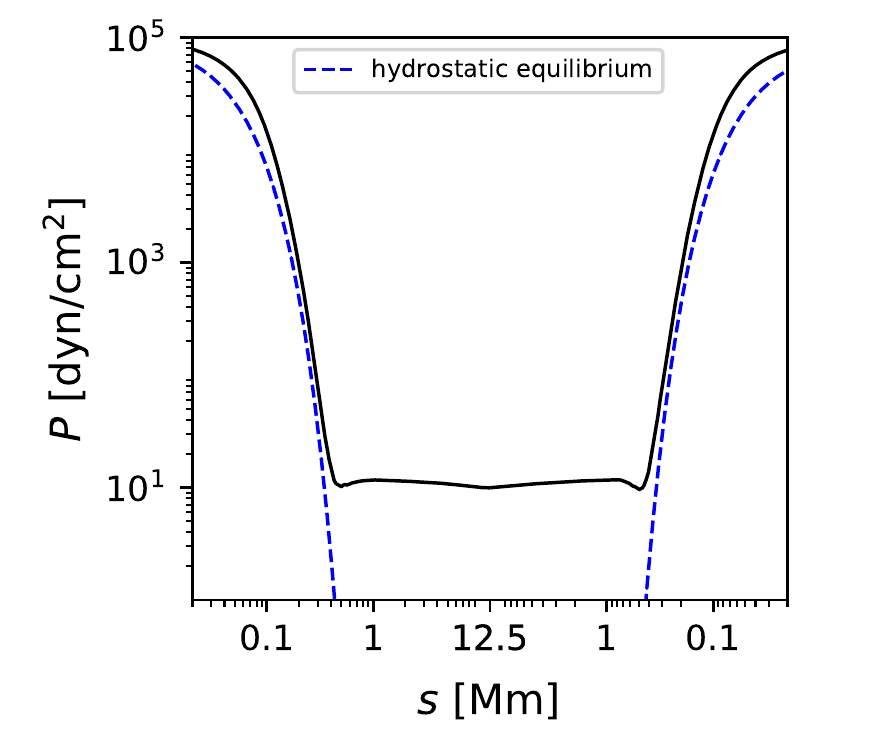}
\end{centering}
  \caption{Comparison of the time-averaged profile of the gas pressure for $l_{\rm loop}=12.5$ Mm and $T_{\rm off}=2000$ K (black solid line) to the isothermal hydrostatic profile for $T=T_{\rm ch}=1500$ K (blue dashed line). }
  \label{fig9}
\end{figure} 

When the loop is short ($l_{\rm loop}=12.5$ Mm), a significant difference is seen in $L_{\rm X}$ for different $T_{\rm min}$ cases.
$L_{\rm x}$ for $T_{\rm min}< 2000$ K is nearly 2 orders of magnitude higher than that for $T_{\rm min}\sim 4000$ K because in high-$T_{\rm min}$ cases, the volume with $>1.1\times 10^6$ K is severely reduced (Figure \ref{fig5}).
The data points for $T_{\rm off}=6000$ and $6500$ K do not appear in the displayed range of both $L_{\rm X}$ and $L_{\rm EUV}$ because the gas is not heated up to $10^5$ K (Figures \ref{fig4} and \ref{fig5}).
The hot region emitting X-rays is small in the short loop, so that $L_{\rm X}$ in $T_{\rm min}<4000$ K is also weaker than that in longer loops.
    
\section{Discussion}
\label{sec4}

\subsection{Condition for the Formation of the Corona}
\label{sec41}

As seen in Section \ref{sec3}, short loops with high $T_{\rm min}$ fail to produce the million-Kelvin corona. 
We here quantitatively discuss the condition of the coronal formation in terms of $T_{\rm min}$ and $l_{\rm loop}$, from the energetics in the loop.

We first define the pressure scale height ($H_{\rm ch}$) in the chromosphere under the isothermal approximation, 
\begin{align}
& H_{\rm ch}= \frac{k_B}{g \mu m_H}  \times T_{\rm ch} ,
\label{eq18}
\end{align}
where $\mu =1.2$ as a typical value for the weakly ionized gas in the photosphere and chromosphere \citep{Cranmer2011, Zurbriggen2020}.
$T_{\rm ch}$ is the temperature at the characteristic density $\rho_{\rm ch}=10^{-9}$ g cm$^{-3}$ in the low chromosphere.
We note that our simulations give $T_{\rm min} < T_{\rm ch} < T_{\rm off}$. 

\begin{figure}[t]
\begin{centering}
\includegraphics{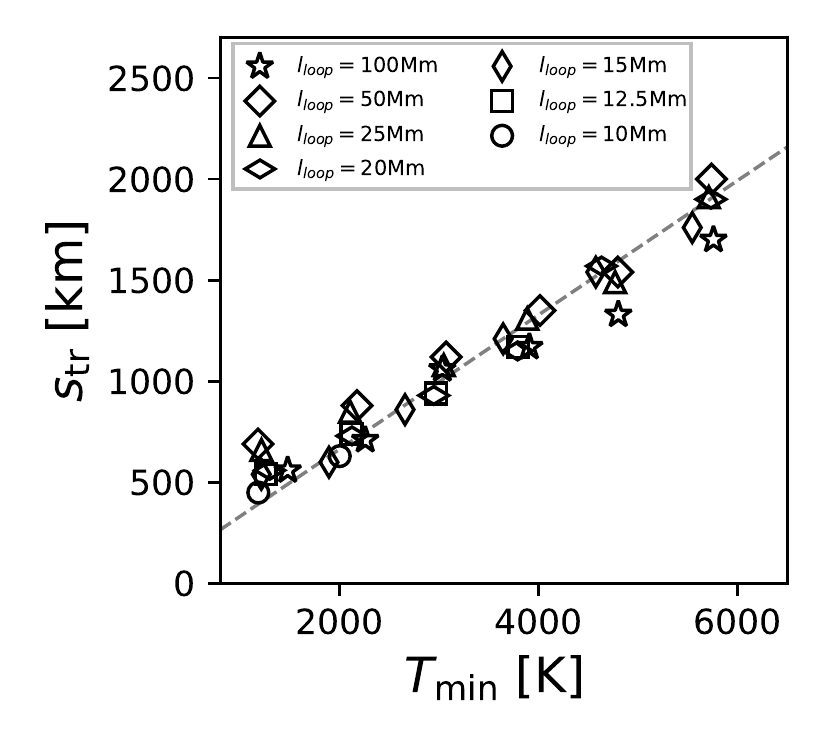}
\end{centering}
  \caption{$s_{\rm tr}$ with $T_{\rm min}$ for the cases giving $T_{\rm top}>0.5$ MK. The dashed line is the fitting line given by Equation (\ref{eq20}).}
  \label{fig10}
\end{figure} 

We define $s_{\rm tr}$ as the distance between the photosphere and the transition region (where $T=10^5$ K). 
Utilizing Equation (\ref{eq18}), we can estimate 
\begin{align}
\begin{split}
s_{\rm tr}&=H_{\rm ch} \times \mathrm{ln}\left(\frac{P_{\rm ph}}{P_{\rm tr}}\right),
\label{eq19}
\end{split}
\end{align}
where $P_{\rm ph}$ and $P_{\rm tr}$ are the pressures at the photosphere and the transition region with $T=10^5$ K.
Figure \ref{fig8} shows the average value of $P_{\rm tr} \sim 10$ dyn cm$^{-2}$ with a weak dependence on $T_{\rm min}$. 

In deriving Equation (\ref{eq19}), we simply assume that the gas is isothermal with $T = T_{\rm ch}$ throughout the chromosphere.
To validate this assumption, Figure \ref{fig9} compares the profile of the gas pressure from the simulation with $T_{\rm off}=2000$ K (black solid line) and $l_{\rm loop}=12.5$ Mm to that in the hydrodynamic equilibrium with constant temperature $T=T_{\rm ch}=1500$ K (blue dashed line). 
The good agreement between the two lines below the transition region indicates that the isothermal approximation is reasonable to estimate $s_{\rm tr}$. 


$T_{\rm ch}$ is proportional to $T_{\rm min}$ in our simulations; we obtain the average trend of $T_{\rm ch}=1.1 T_{\rm min}$. 
Then, from Equations (\ref{eq18}) and (\ref{eq19}), $s_{\rm tr}$ has a linear dependence on $T_{\rm min}$. 
Figure \ref{fig10} shows the relation between $s_{\rm tr}$ and $T_{\rm min}$ for the cases that give the temperature at the top $T_{\rm top} > 0.5$ MK.
The dashed line shows a linear fit:  
\begin{align}
&s_{\rm tr}= 0.33 \left( \frac{T_{\rm min}}{10^3 \hspace{1mm}\rm K} \right) \hspace{2mm} \rm{Mm}.
\label{eq20}
\end{align}
When we directly derive the relation using $T_{\rm ch}=1.1 T_{\rm min}$, $P_{\rm ph}=10^5$ dyn cm$^{-2}$ and $P_{\rm tr}=10$ dyn cm$^{-2}$, Equation (\ref{eq19}) yields $s_{\rm tr}=0.25 (T_{\rm min}/10^3$ $\rm K)$, which roughly reproduces Equation (\ref{eq20}).

\begin{figure}[t]
\begin{centering}
\includegraphics{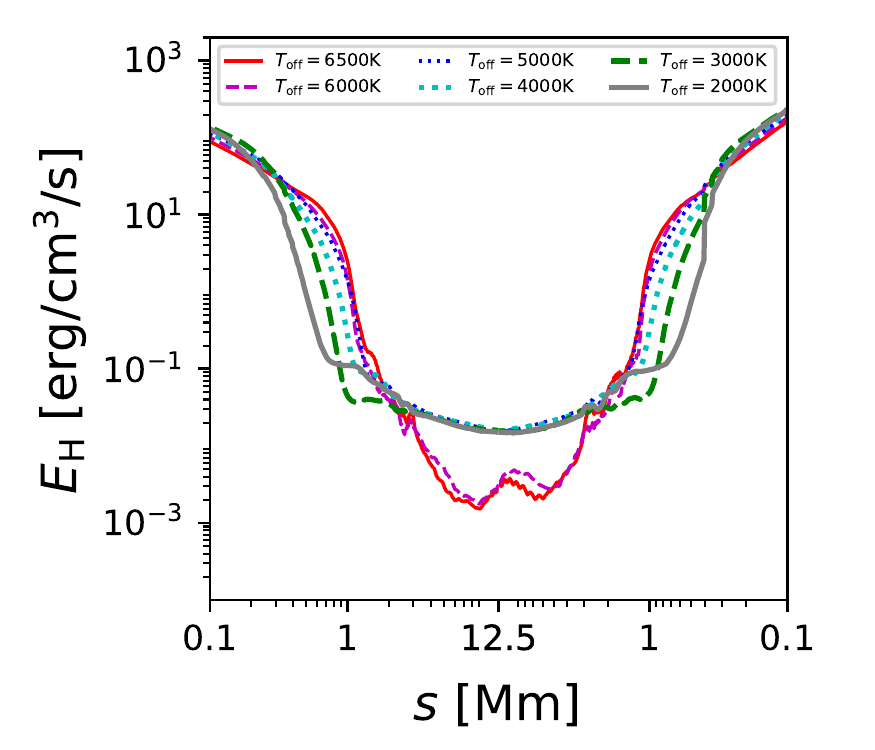}
\end{centering}
  \caption{Volumetric heating rates for $l_{\rm loop}=12.5$ Mm  with different $T_{\rm off}$. }
  \label{fig11}
\end{figure} 

To inspect the energetics in loops, we evaluate the volumetric heating rate by the dissipation of Alfv\'{e}n waves, 
\begin{align}
E_H = -\nabla \cdot \left[ -\frac{1}{4\pi} B_s (\bm{B_{\perp}}\cdot\bm{v_{\perp}})+ \left(\frac{1}{2} \rho v_{\perp}^2 + \frac{B_{\perp}^2}{4\pi}\right)v_s \right].
\label{eq21}
\end{align}
Figure \ref{fig11} compares the profiles of $E_H$ for $l_{\rm loop}=12.5$ Mm. 
In the cases of $T_{\rm off} \geq 6000$ K, $E_H$ drops with the height up to the loop top, while low-$T_{\rm off}$ cases show the uniform heating rate in the corona.
Here, we define 
\begin{align}
l_{\rm cor} = l_{\rm loop} - s_{\rm tr} 
\label{eq22}
\end{align}
as the half length of the coronal region. 
Even though $l_{\rm loop}$ is the same, $l_{\rm cor}$ is smaller for higher $T_{\rm off}$, which gives a larger downward conductive flux ($\propto l_{\rm cor}^{-1}$) for a fixed $T_{\rm top}$.
Hence, a larger $E_H$ is required to maintain the corona for higher $T_{\rm off}$ cases.
However, Figure \ref{fig11} indicates that this is not satisfied, and therefore, the hot corona with $T_{\rm top} > 0.5$ MK is not formed in the cases with $T_{\rm off}\ge 6000$ K (Figures \ref{fig4} and \ref{fig5} ). 

\begin{figure}[t]
\begin{centering}
\includegraphics{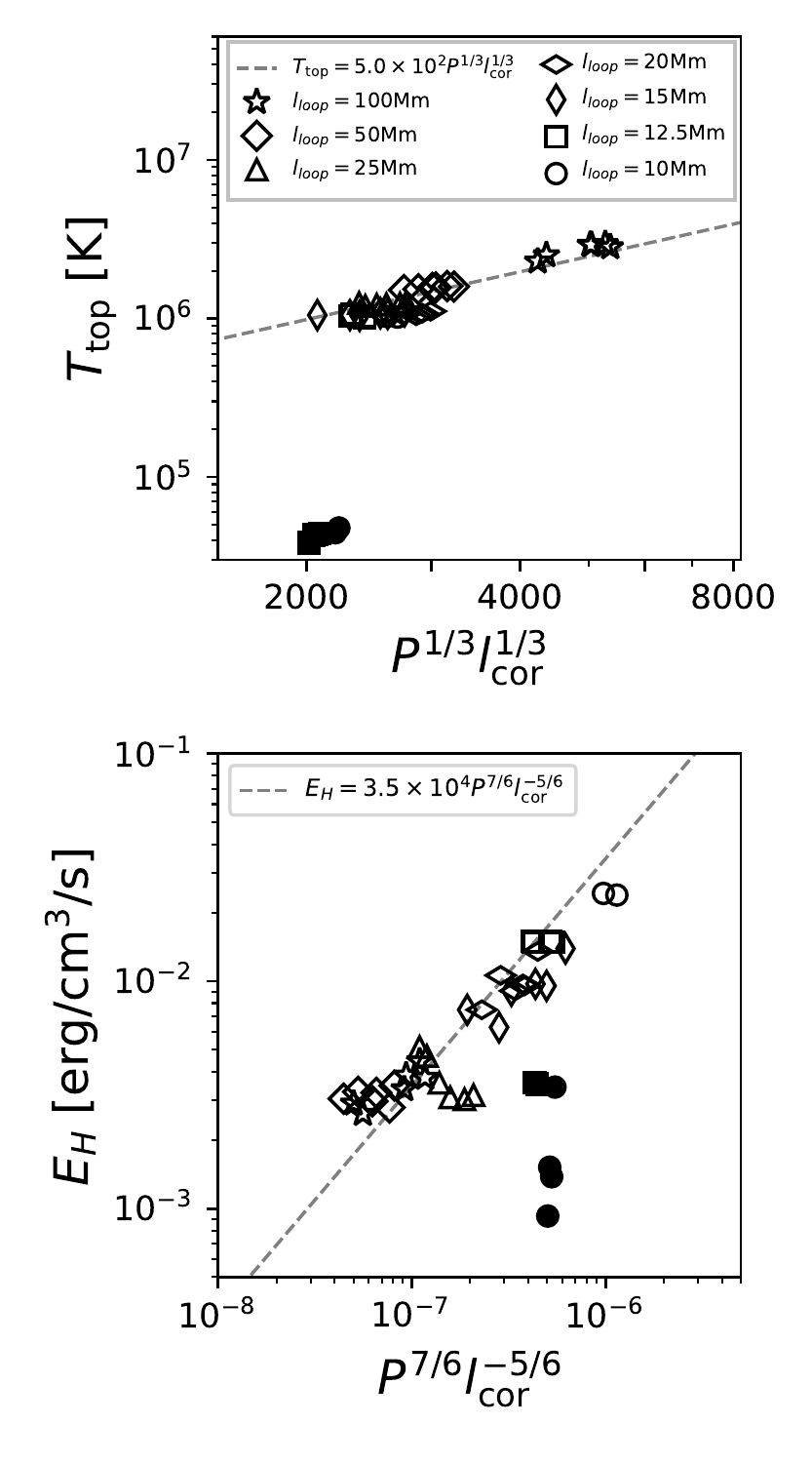}
\end{centering}
  \caption{$T_{\rm top}$ with $P^{1/3}l_{\rm cor}^{1/3}$ (top) and $E_H$ with $P^{7/6} l_{\rm cor}^{-5/6}$ (bottom) for all the simulated cases. Open symbols correspond to the cases in which the corona is formed ($T_{\rm top}>0.5$ MK).
  Filled symbols are the cases in which the corona is not formed ($T_{\rm top}\leq 0.5$ MK). The dashed lines are the fitting lines of Equations (\ref{eq25}) and (\ref{eq26}).}
  \label{fig12}
\end{figure}

For static coronal loops, the maximum temperature and the volumetric heating rate are related to $l_{\rm cor}$ (Equation (\ref{eq22})) and the pressure, $P$, in a loop via the RTV scaling laws \citep{Rosner1978, Serio1981, Zhuleku2020}:
\begin{align}
&T_{\rm RTV} \mathrm{[K]} \simeq 1.4\times 10^3 \left(\frac{P}{\rm dyn \hspace{1mm}cm^{-2}}\right)^{1/3} \left(\frac{l_{\rm cor}}{\rm cm}\right)^{1/3},
\label{eq23}
\end{align}
\begin{align}
\begin{split}
&E_{H_{\rm RTV}} \mathrm{[erg \hspace{1mm} cm}^{-3} s^{-1}]  \\
&\simeq 9.8 \times 10^4 \left(\frac{P}{\rm dyn \hspace{1mm}cm^{-2}}\right)^{7/6} \left(\frac{l_{\rm cor}}{\rm cm}\right)^{-5/6}.
\label{eq24}
\end{split}
\end{align}
We here compare our results with Equations (\ref{eq23}) and (\ref{eq24}).
Figure \ref{fig12} shows $T_{\rm top}$ and $E_H$ with $P^{1/3} l_{\rm cor}^{1/3}$ and $P^{7/6} l_{\rm cor}^{-5/6}$ for all the simulated cases.
We use the value of $P_{\rm tr}$ for $P$ and the averaged value over 8 Mm around the loop top for $E_H$.
For $s_{\rm tr}$, which is necessary to determine $l_{\rm cor}$ (Equation (\ref{eq22})), we adopt the values obtained from our simulations for $T_{\rm top}>0.5$ MK (open symbols) and the values expected from Equation (\ref{eq20}) for $T_{\rm top}\leq 0.5$ MK (filled symbols).

\begin{figure}[t]
\begin{centering}
\includegraphics{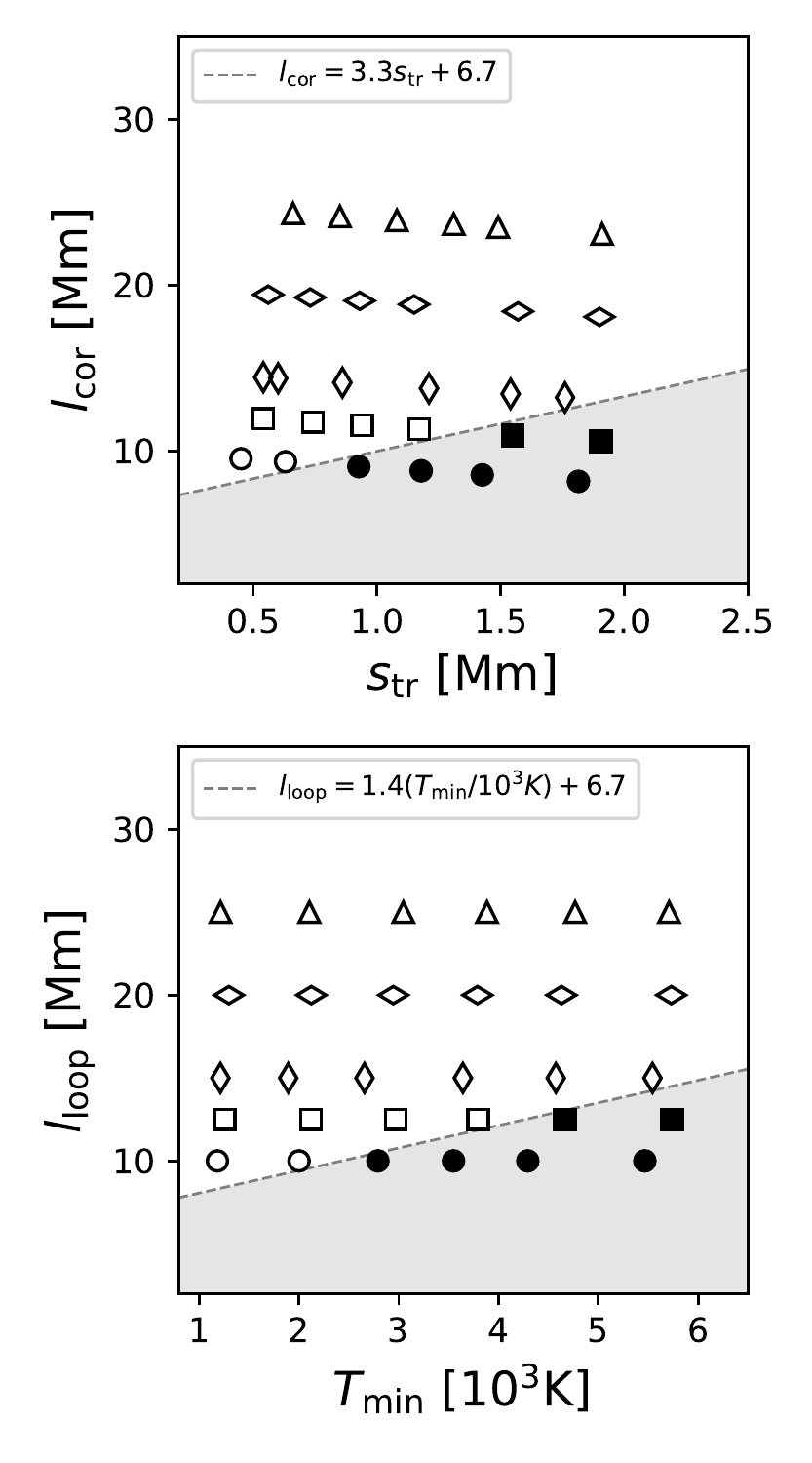}
\end{centering}
  \caption{$l_{\rm cor}$ with $s_{\rm tr}$ (top) and $l_{\rm loop}$ with $T_{\rm min}$ (bottom).  
  The symbols are the same as those in Figure \ref{fig12}.
  The hot corona is not formed in the shaded region where Equations (\ref{eq27}) and (\ref{eq28}) are not satisfied. }
   \label{fig13}
\end{figure} 

$T_{\rm top}$ and $E_H$ of the cases with the high-temperature corona (open symbols) exhibit the same dependencies as the original RTV scaling laws.
The fitting to the open symbols gives
\begin{align}
&T_{\rm top} \simeq 5.0 \times 10^2 \left(\frac{P}{\rm dyn \hspace{1mm}cm^{-2}}\right)^{1/3} \left(\frac{l_{\rm cor}}{\rm cm}\right)^{1/3},
\label{eq25}
\end{align}
\begin{align}
&E_H \simeq 3.5 \times 10^4 \left(\frac{P}{\rm dyn \hspace{1mm}cm^{-2}}\right)^{7/6}\left(\frac{l_{\rm cor}}{\rm cm}\right)^{-5/6}.
\label{eq26}
\end{align}
The deviations from Equations (\ref{eq23}) and (\ref{eq24}) probably come from the lack of numerical resolution at the transition region (see the Appendix of \cite{Shoda2021}). 
On the other hand, all the filled symbols ($T_{\rm top}\leq 0.5$ MK) are below the fitting lines of Equations (\ref{eq25}) and (\ref{eq26}); because of the insufficient heating ($E_H$ of the bottom panel), the upper layer cannot be heated up to the sufficiently high temperature (top panel). 

To clarify the condition for the formation of the high-temperature corona, we summarize the cases with (without) the corona of $T_{\rm top} > 0.5$ MK by open (filled) symbols in a $s_{\rm tr}-l_{\rm cor}$ plane (top panel of Figure \ref{fig13}).
We find that the corona is formed when
\begin{align}
&l_{\rm cor} > 3.3 s_{\rm tr} + 6.7 \hspace{2mm} \rm{Mm}.
\label{eq27}
\end{align}
Using Equations (\ref{eq20}) and (\ref{eq22}), we can further rewrite Equation (\ref{eq27}) with $l_{\rm loop}$ and $T_{\rm min}$;
\begin{align}
l_{\rm loop} > 1.4 \left(\frac{T_{\rm min}}{10^3 \hspace{1mm}\rm K}\right) + 6.7 \hspace{2mm} \rm{Mm}.
\label{eq28}
\end{align}
When we directly fit the data in a $T_{\rm min}-l_{\rm loop}$ plane, we obtain the same condition of the coronal formation (bottom panel of Figure \ref{fig13}).
For loops with small $l_{\rm loop}$ and high $T_{\rm min}$, Equation (\ref{eq28}) does not hold; the corona with $T > 0.5$ MK is not formed in the shaded region.
It quantitatively confirms the results in Section \ref{sec31} that the coronal properties of the short loops ($l_{\rm loop}\leq 12.5$ Mm) are determined by the temperature and density structures of the chromosphere, while those of the long loops are hardly affected.

Equation (\ref{eq28}) indicates that at least $l_{\rm loop} > 6.7$ Mm is required for the formation of the corona in the limit of $T_{\rm min} \rightarrow 0$ and that a small increase in the chromospheric temperature drastically enhances the required $l_{\rm loop}$. 
Specifically, when $s_{\rm tr}$ increases by 1 Mm, $l_{\rm loop}$ needs to increase by 4.3 Mm to form the corona (see Equations (\ref{eq22}) and (\ref{eq27})).
This is because a large amount of Alfv\'{e}n waves dissipates in the extended chromosphere when $s_{\rm tr}$ increases. 
As a result, a smaller fraction of the input Alfv\'{e}nic Poynting flux survives in the upper layer, leading to the loss of the corona.
We also discuss the importance of the wave dissipation in the chromosphere in Section \ref{sec42}.

\begin{figure}[!t]
\begin{centering}
\includegraphics{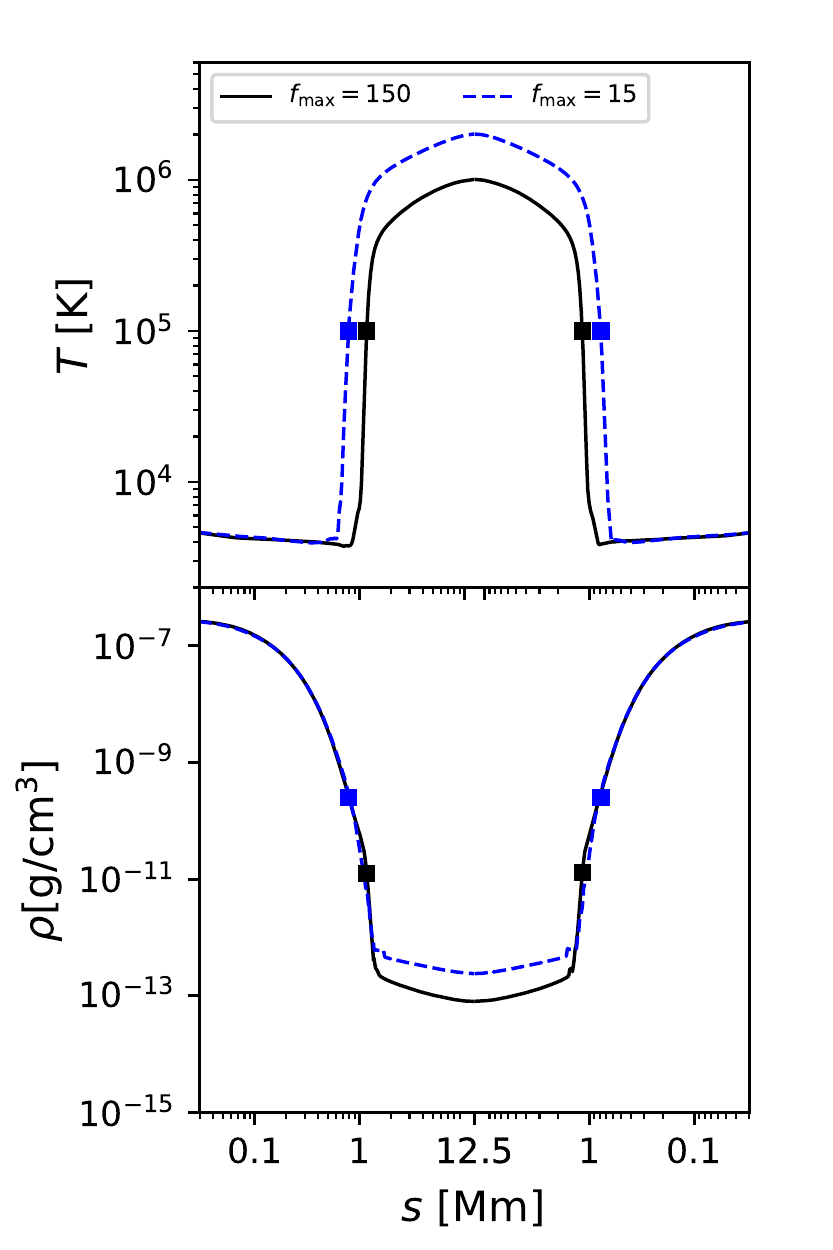}
\end{centering}
  \caption{Time-averaged loop profiles of the temperature (top) and density (bottom) for $f_{\rm max}=150$ (black solid line) and $15$ (blue dashed line) with $l_{\rm loop}=12.5$ Mm and $T_{\rm off}=5000$ K. }
  \label{fig14}
\end{figure} 
\begin{figure}[t]
\begin{centering}
\includegraphics{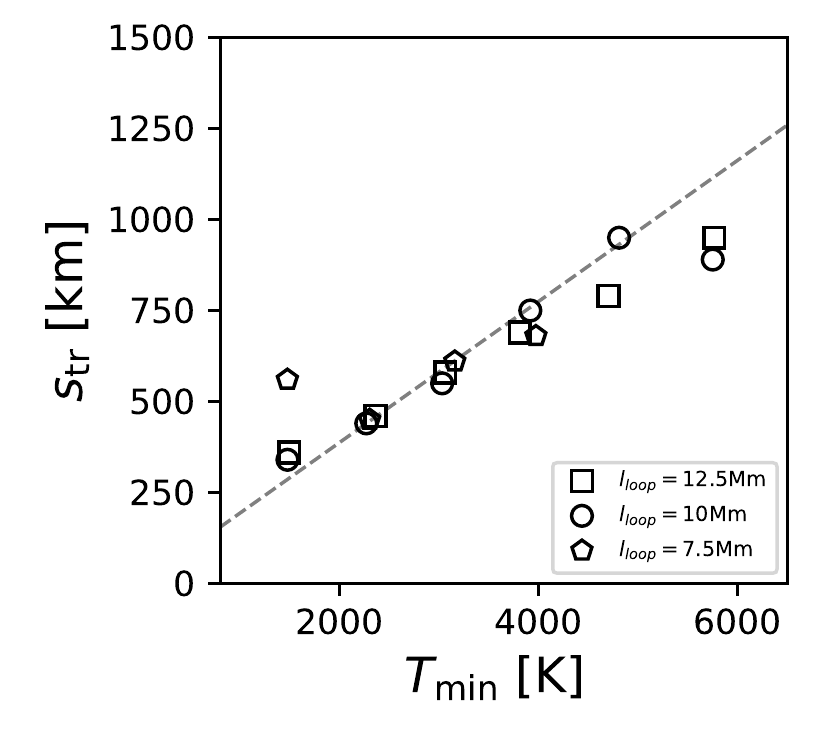}
\end{centering}
  \caption{$s_{\rm tr}$ with $T_{\rm min}$ for the cases of $f_{\rm max}=15$. The dashed line is the fitting line given by Equation (\ref{eq29}).}
  \label{fig15}
\end{figure}

\begin{figure}[!t]
\begin{centering}
\includegraphics{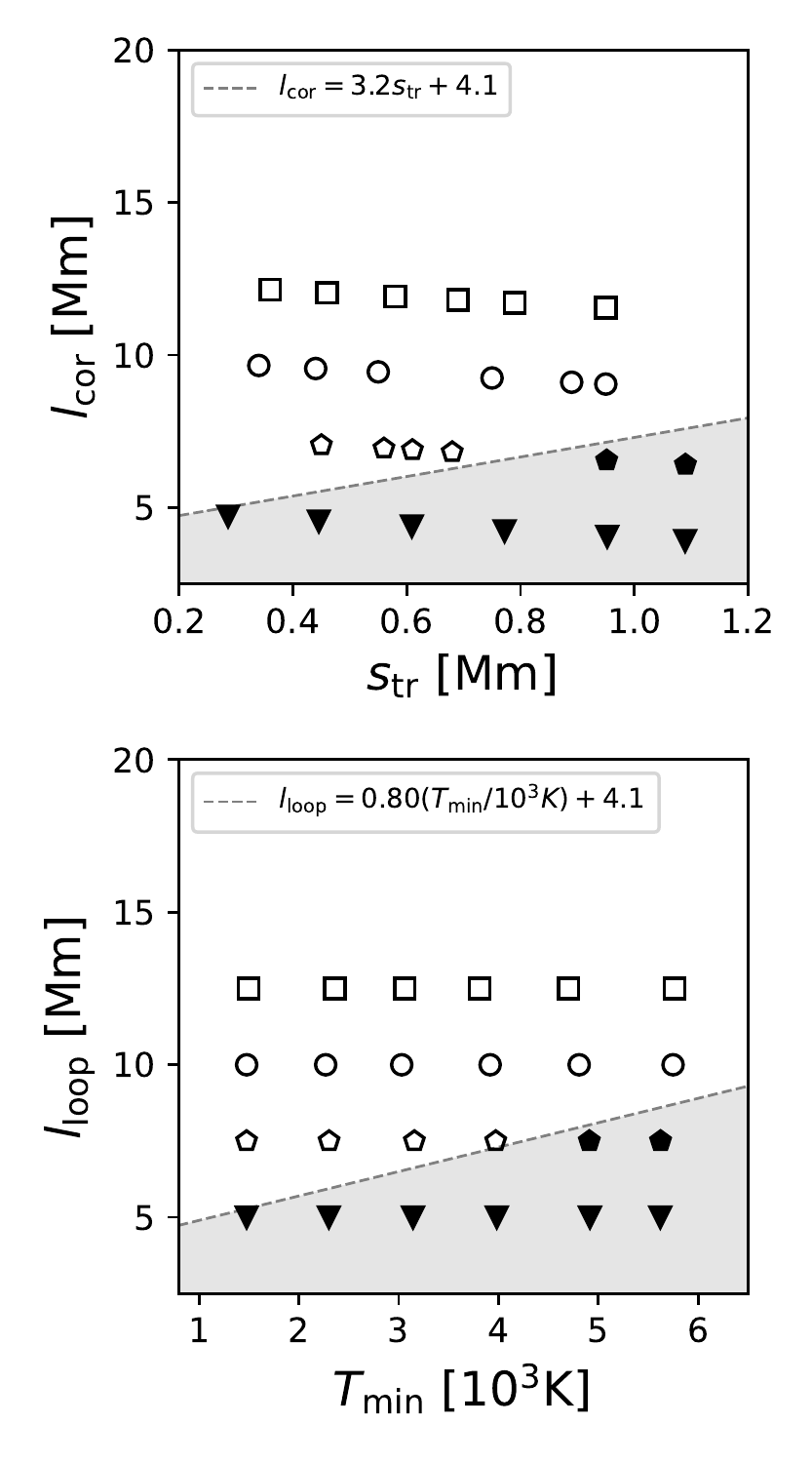}
\end{centering}
  \caption{$l_{\rm cor}$ with $s_{\rm tr}$ (top) and $l_{\rm loop}$ with $T_{\rm min}$ (bottom) for $f_{\rm max}=15$. The hot corona is not formed in the shaded region where Equations (\ref{eq30}) and (\ref{eq31}) are not satisfied.}
  \label{fig16}
\end{figure} 

\subsection{Condition for Active Regions}
\label{sec43}
In the former sections, we have simulated the loops with the coronal field strength $B_{\rm cor}=10.5$ G by setting $f_{\rm max}=150$ (Figure \ref{fig1}), focusing on magnetic loops in the quiet Sun. 
The million-Kelvin corona is not formed in loops with $l_{\rm loop} \leq 12.5$ Mm and hot chromosphere of $T_{\rm min}>4000$ K. 
In reality, however, bright short loops have been observed in solar active regions \citep{Aschwanden2008}, which is apparently contradictory to our conclusion. 
This discrepancy occurs possibly because the coronal-formation condition is affected by the coronal field strength.

In this section, we change the numerical setup to derive the condition for the coronal formation with a stronger magnetic field.
We here test the cases with $B_{\rm cor}=105$ G by setting $f_{\rm max}=15$. 
We note that $B_{\rm ph}$ is unchanged in the active-region setting, so that the equipartition on the photosphere is still satisfied (the black solid lines in Figure \ref{fig1}).
To find the condition for the formation of the corona, we carry out the simulations particularly in short loops with $l_{\rm loop}=5.0, 7.5, 10,$ and  $12.5$ Mm.

The coronal property is affected by $f_{\rm max}$. Figure \ref{fig14} compares the time-averaged profiles for $f_{\rm max}=150$ and $15$ with $l_{\rm loop}=12.5$ Mm and $T_{\rm off}=5000$ K.
It is clearly seen that smaller $f_{\rm max}$, namely, a stronger coronal magnetic field, gives a higher coronal temperature and density.

The black and blue squares in Figure \ref{fig14} show that the location of the transition region ($T=10^5$ K) also depends on $f_{\rm max}$.
In the smaller-$f_{\rm max}$ case, the conduction-radiation balance is achieved at a lower height owing to the enhanced heating.
The density at the transition region is also higher because the large thermal conduction leads to the efficient chromospheric evaporation.
Figure \ref{fig15} shows the relation between $s_{\rm tr}$ and $T_{\rm min}$ for $f_{\rm max}=15$. 
The linear fitting gives
\begin{align}
&s_{\rm tr}= 0.19 \left( \frac{T_{\rm min}}{10^3 \hspace{1mm}\rm K} \right) \hspace{2mm} \rm{Mm}.
\label{eq29}
\end{align}

We plot the simulated cases in $s_{\rm tr}-l_{\rm cor}$ and $T_{\rm min}-l_{\rm loop}$ planes in Figure \ref{fig16}, where we estimate $s_{\rm tr}$ for $T_{\rm top}<0.5$ MK (filled symbols) by using Equation (\ref{eq29}).
Compared to the cases with weak magnetic fields (Figure \ref{fig13}), the corona can be formed in smaller $l_{\rm loop}$.
The conditions for the coronal formation are obtained as follows;
\begin{align}
&l_{\rm cor} > 3.2 s_{\rm tr} + 4.1 \hspace{2mm} \rm{Mm},
\label{eq30}
\end{align}
\begin{align}
&l_{\rm loop} > 0.80 \left(\frac{T_{\rm min}}{10^3 \hspace{1mm}\rm K}\right) + 4.1 \hspace{2mm} \rm{Mm}. 
\label{eq31}
\end{align}
Comparing Equations (\ref{eq28}) and (\ref{eq31}), it turns out that the coronal field strength highly affects the condition of forming the corona.
The stronger coronal magnetic field allows for the formation of the hot corona in shorter loops even though the chromospheric temperature is high.
The dependence on $T_{\rm min}$ is weaker because the height of the transition region is lower for stronger coronal fields (Equations (\ref{eq20}) and (\ref{eq29})).

\subsection{Effect of the Photospheric Perturbations}
\label{sec44}

In this study, the simulations are carried out with the typical setup for the wave injection from the photosphere (Section \ref{sec25}). 
The physical properties of the photospheric driver affect the coronal heating, and accordingly the condition for the coronal formation.

The amplitude of velocity perturbation has a certain effect on the coronal properties \citep{Antolin2010, Suzuki2013}.
When the amplitude is increased, the heating is enhanced owing to the large input Poynting flux.
The constraints of the coronal formation by the chromospheric temperature is then expected to be loosened.

In addition, the heating efficiency from Alfv\'{e}n waves is controlled by the range of the frequency spectrum, which determines the ratio of high/low-frequency components of the injected waves.
We will also discuss the different heating mechanisms in the next subsection.

\subsection{Heating Mechanisms}
\label{sec46}

High-frequency waves with the wavelength smaller than the loop length behave as propagating waves. 
They dissipate via turbulent cascade and nonlinear mode conversion.
The high-frequency longitudinal fluctuations, as well as the transverse ones, also play a role in heating from mode conversion \citep{Shimizu2022}. 
In contrast, the low-frequency waves with wavelengths larger than the loop length have the different behaviors. 
A moderate fraction of the low-frequency fluctuations in the corona is trapped between the transition regions due to reflection. 
Thus, these fluctuations gradually evolve into turbulent state due to field-line braiding.
They eventually contribute to the heating by magnetic reconnection of braiding field lines \citep{Parker1972, Parker1983}, which can be handled in the diffusion terms in Equations (\ref{eq7}) and (\ref{eq8}).

When we compare the total amount of the heating for loops of different lengths, it is larger for longer loops because they have a larger spatial region, and then a larger fraction of the Poynting flux entering the corona can dissipate to heat the gas.
In particular, the decay of the Poynting flux in the low-density region at a high altitude gives a large heating rate per mass, which allows the gas to be heated to high temperatures.
On the other hand, short loops do not have a sufficient length for the energy to decay, and therefore a small fraction of the Poynting flux that dissipates into heat.
Therefore, some cases fail to produce a hot corona due to the insufficient heating (filled symbols in the bottom panel of Figure \ref{fig12}).


Recently, localized transient brightenings, "campfires", were detected in the quiet Sun with high-resolution EUV observations with the Extreme Ultraviolet Imager on board Solar Orbiter.
They are generally short-lived with $10 - 200$ sec and show small loop-like structures with the full length of $0.4 - 4$ Mm \citep{Berghmans2021}.
This loop length does not satisfy Equations (\ref{eq28}) and (\ref{eq31}).
The inconsistency is probably because these transient brightenings are originated from impulsive magnetic reconnections that are associated with interactions between neighboring loops \citep{Chen2021}.
Our simulations cannot directly handle this phenomenon due to the limited treatment of the time-steady heating in a single isolated loop.
Generalizing Equations (\ref{eq28}) and (\ref{eq31}) for various heating mechanisms is a direction of our future studies.

\subsection{Numerical Resolution for the Transition Region}
\label{sec48}

In our simulation, the transition region is resolved with the grid size of 5 km for all the cases (Section \ref{sec26}).
However, particularly when the hot corona forms in short loops, higher resolution may be required to accurately solve the interaction between the downward conductive flux from the corona and the chromospheric evaporation.
Previous numerical studies have reported that the coronal density is underestimated when the transition region is not sufficiently resolved \citep{Bradshaw2013, Johnston2017}.

To inspect the effect of the numerical resolution, we run the additional simulations with a finer grid size of 2 km at the transition region.
For the cases of $B_{\rm cor}=10.5$ G with $l_{\rm loop}=12.5$ Mm, $T_{\rm top}$ and $\rho_{\rm top}$ are plotted in Figure \ref{fig4} (cyan squares). 
When the cyan squares are compared with the blue squares, $\rho_{\rm top}$ is slightly increased for high-resolution cases, while $T_{\rm top}$ shows almost the same values.
The same trend is also seen for the cases of $B_{\rm cor}=105$ G with $l_{\rm loop}=7.5$ Mm.
These indicate that the condition of the coronal formation (Equations (\ref{eq28}) and (\ref{eq31})) is not sensitively affected by the spatial resolution.


\subsection{Radiative Loss in the Chromosphere}
\label{sec47}

To perform simulations with a variety of thermal structures, we artificially set the chromospheric temperature by controlling the radiative cooling.
Our simulations employ the simple empirical function for the radiative loss in the chromosphere, following (\cite{Anderson1989}; Equation (\ref{eq16})). 
As our purpose is to characterize the properties of the upper atmosphere by $T_{\rm min}$, the precise modeling of the radiative loss is out of the scope of this study.

We here discuss the impact of the different radiative loss function on our results.
While Equation (\ref{eq16}) assumes a constant cooling rate per mass, the radiative loss in the chromosphere is in fact dominated by several strong spectral lines formed at different locations.
Currently, a sophisticated recipe for radiative cooling and heating has been developed from the detailed radiative transfer calculation to reproduce the spectral features in the chromosphere \citep{Carlsson2012}.
\cite{Wang2020} have compared the resultant atmospheres using the different radiative loss functions by \cite{Anderson1989} and \cite{Carlsson2012}.
They have reported that, although the detailed treatment of \cite{Carlsson2012} is more appropriate to study dynamical phenomena in the chromosphere, the time-averaged radiative losses and temperature profiles are in close agreement between the two treatments.
Therefore, the different radiative loss function does not probably affect the condition for the formation of the quasi-steady corona (Equations (\ref{eq28}) and (\ref{eq31})). 


\begin{figure}[!t]
\begin{centering}
\includegraphics{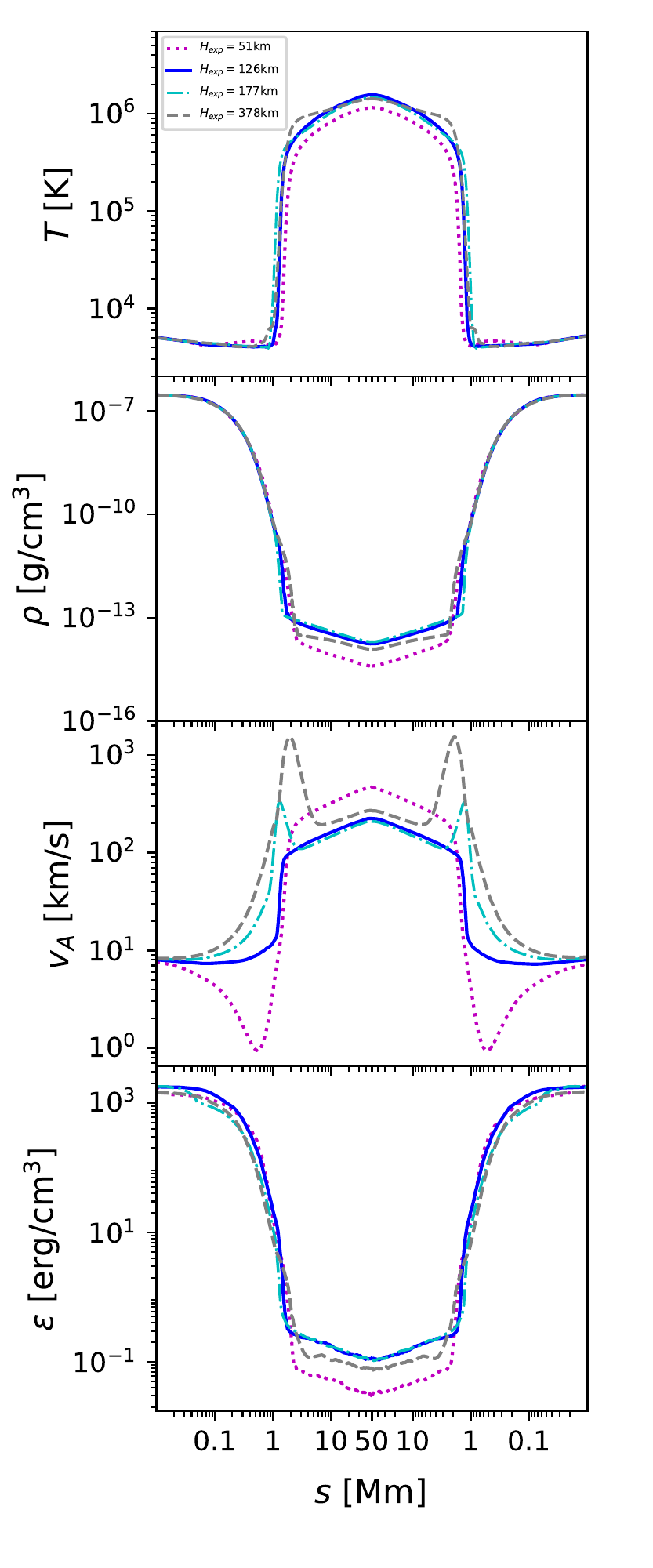}
\end{centering}
  \caption{Time-averaged loop profiles of temperature, density, Alfv\'{e}n speed, and energy density of the Alfv\'{e}n waves for different $H_{\rm exp}$ when $l_{\rm loop}=50$ Mm and $T_{\rm off}=5000$ K.}
  \label{fig17}
\end{figure} 

\subsection{Geometry of the Flux Tube}
\label{sec42}

So far, we have fixed the geometry of the flux tubes based on the scale height in the chromosphere (Section \ref{sec22}).
In reality, however, the chromosphere is expected to exhibit a variety of  magnetic structures.
Here we study the effect of the flux-tube configuration on the coronal properties. 

We run the simulations with different profiles of $f(s)$ shown in Figure \ref{fig1}; we vary the value of $H_{\rm exp}$ while keeping $f_{\rm max}=150$ and $T_{\rm off} = 5000$ K.  
The simulations for $T_{\rm off}=5000$ K in Section \ref{sec3} adopt $H_{\rm exp}=126$ km (blue line). 
We here set $H_{\rm exp}=51$ km, 177 km, and $378$ km in Equations (\ref{eq3}) and (\ref{eq4}), and in these cases $f(s)$ reaches $f_{\rm max}$ at 1.1 Mm, 3.0 Mm, and 7.0 Mm, respectively. 

\begin{figure}[t]
\begin{centering}
\includegraphics{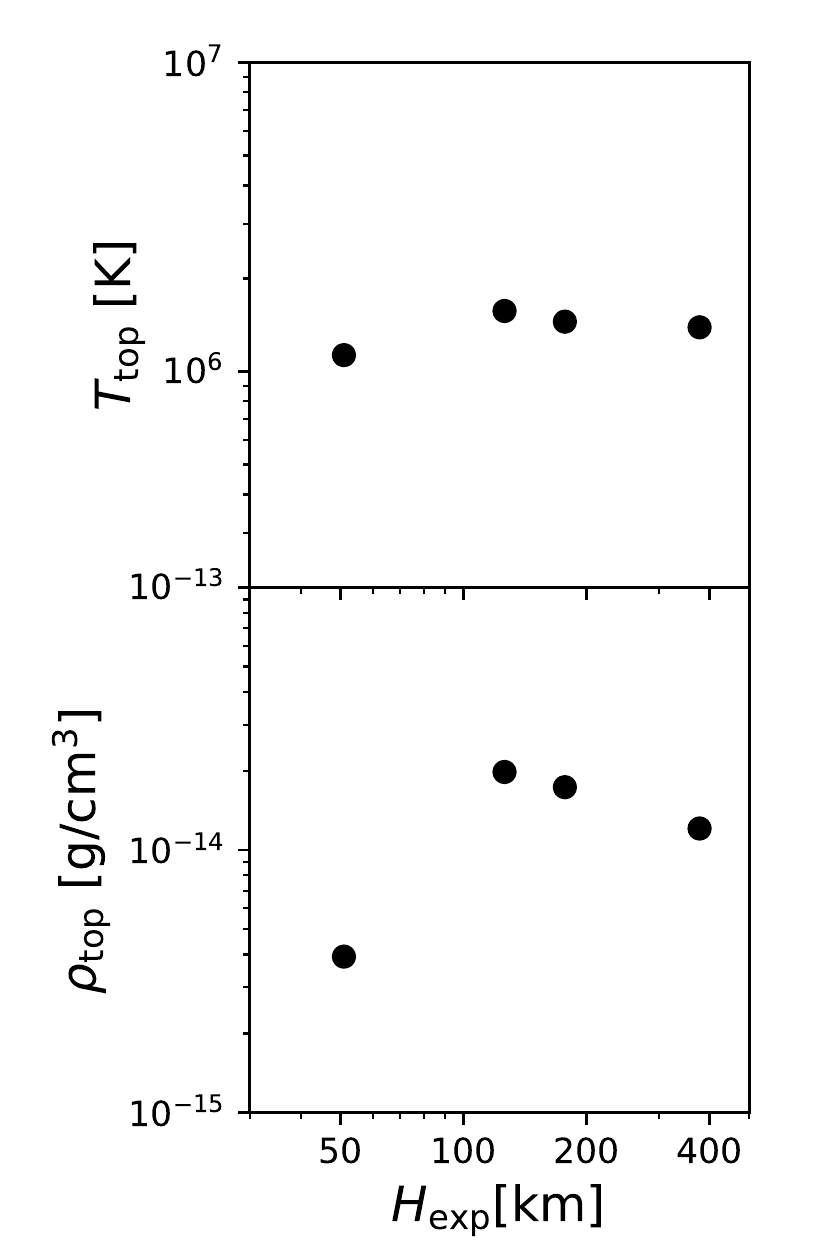}
\end{centering}
  \caption{Temperature and density at the loop top with $H_{\rm exp}$ when $l_{\rm loop}=50$ Mm and $T_{\rm off}=5000$ K.}
  \label{fig18}
\end{figure} 
\begin{figure}[htb]
\begin{centering}
\includegraphics{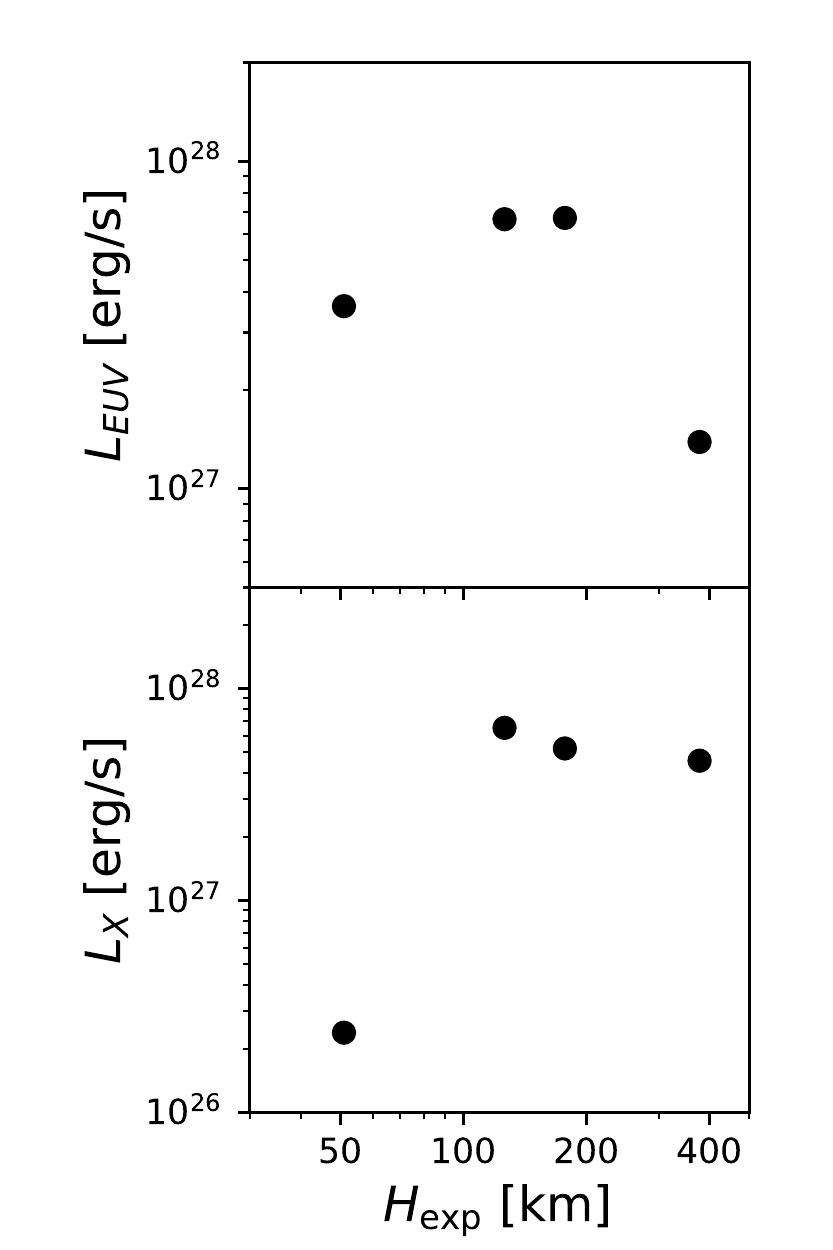}
\end{centering}
  \caption{$L_{\rm EUV}$ and $L_{\rm X}$ with $H_{\rm exp}$ when $l_{\rm loop}=50$ Mm and $T_{\rm off}=5000$ K. }
  \label{fig19}
\end{figure} 

The loop profiles are compared in Figure \ref{fig17} and $T_{\rm top}$ and $\rho_{\rm top}$ are shown in Figure \ref{fig18}. 
The corona is largely affected by the location of the expansion because it affects the propagation of Alfv\'{e}n waves through the change in $v_{\rm A}$. 
The third panel of Figure \ref{fig17} shows that the profiles of $v_{\rm A}$ are quite different. 
For example, in the case of $H_{\rm exp}=51$ km, the Alfv\'{e}n speed first decreases in height below the transition region because the magnetic field, $B(s)=B_{\rm ph}/f(s)$, quickly decreases due to the rapid expansion of the  flux tube. 
The variation in $v_{\rm A}$ causes the reflection of Alfv\'{e}n waves back to the photosphere \citep{An1990, Velli1993, Cranmer2005}. 

To see this effect, we present the energy density of Alfv\'{e}n waves,
\begin{align}
& \epsilon = \frac{1}{2} \rho v_{\perp}^2 + \frac{B_{\perp} ^2}{8\pi},
\label{eq32}
\end{align}
in the bottom panel of Figure \ref{fig17}.
The case of $H_{\rm exp} = 51$ km gives the smallest survived Alfv\'{e}nic energy density in the corona mainly because of the reflection below the coronal base.  

The case of $H_{\rm exp}=378$ km also shows the relatively cool corona. The Alfv\'{e}n speed has a peak at the coronal base because of the high position of the expansion, contrary to the case of $H_{\rm exp}=51$ km. 
In the presence of local maximum in the Alfvén speed, only a small fraction of the input Alfvénic wave energy is transmitted into the corona due to reflection. 
This is confirmed in the comparison of $\epsilon$ (bottom panel of Figure \ref{fig17}).  
Our standard case of $H_{\rm exp}=126$ km gives the highest temperature and density in the corona, which indicates that the efficient heating occurs when the Alfv\'{e}n speed smoothly changes without dip nor bump.

In Figure \ref{fig19}, we show the variations in $L_{\rm EUV}$ and $L_{\rm X}$ with respect to $H_{\rm exp}$.
Both are largest at $H_{\exp}=126$ km, which is expected from the results in Figures \ref{fig17} and \ref{fig18}. 
When $H_{\rm exp} = 51$ km, $L_{\rm X}$ is much smaller than the other cases because the coronal density and temperature are small (due to weak heating; see Figure \ref{fig18}).

\section{Summary}
\label{sec5}

We performed the one-dimensional MHD simulations of coronal loops to study the relation between the chromospheric thermal structure and the coronal properties.
The chromospheric temperature, which is characterized by its minimum value $T_{\rm min}$, is controlled by the switch-off temperature of the radiative cooling $T_{\rm off}$. 
We find that the thermal structure of the lower atmosphere certainly has a strong impact on coronal heating.

A common trend obtained in our simulations is that in higher-$T_{\rm min}$ cases, the chromosphere is thicker due to the extended density scale height.
The properties of the upper layer are governed by the loop length in addition to $T_{\rm min}$.

For our standard cases, we assumed the quiet solar atmosphere that exhibits $B_{\rm cor}$ = 10.5 G ($f_{\rm max}=150$).
When $l_{\rm loop} > 12.5$ Mm, the coronal properties are not largely affected by the chromospheric temperature. 
As the gradient of the Alfv\'{e}n speed is smaller for higher-$T_{\rm min}$ cases, a larger fraction of the input Alfv\'{e}nic waves can transmit to the corona, which leads to the formation of a hot and dense corona. 
On the other hand, the dense extended chromosphere with high $T_{\rm min}$ enhances the energy loss by radiative cooling.
The dependence of $T_{\rm top}$ and $\rho_{\rm top}$ on $T_{\rm min}$ is therefore small owing to the cancellation of these two effects.
Correspondingly, most cases show that $L_{\rm EUV}$ and $L_{\rm X}$ give the similar values even for different $T_{\rm min}$. 

However, in short loops with $l_{\rm loop}\leq 12.5$ Mm, it is difficult to form a high-temperature corona in high-$T_{\rm min}$ cases because the fraction of the chromosphere is too large, leaving an insufficient loop length for the corona.
In the cases of $l_{\rm loop}=12.5$ Mm, the corona is not formed when $T_{\rm min}> 4000$ K, while $T_{\rm top}$ and $\rho_{\rm top}$ do not strongly depend on $T_{\rm min}$ when $T_{\rm min}<4000$ K.

From the simulated cases with $B_{\rm cor}$=10.5 G, we have found that the condition for the coronal formation is given by $l_{\rm loop}>1.4(T_{\rm min}/10^3$ $\rm K) + 6.7$ Mm (Equation (\ref{eq28})).
This indicates that a small change in $T_{\rm min}$ gives a large impact on the required $l_{\rm loop}$. 
It is concluded that the coronal heating is inseparably linked to the physical condition of the chromosphere.

In addition, we inspected the condition with the strong coronal magnetic field by setting small $f_{\rm max}$.
When the coronal magnetic field is strong, the hot corona is easily realized even in short loops with hot chromosphere because the strong magnetic field enhances the heating.
The condition is altered to be $l_{\rm loop}>0.8(T_{\rm min}/10^3$ $\rm K) + 4.1$ Mm (Equation (\ref{eq31})) when we adopt a 10 times stronger coronal field strength, 105 G, by reducing $f_{\rm max}$ to 15. 

We also investigated the effect of the geometry of the flux tube on the heating processes by changing the location of the expansion of the flux tube.
Because the shape of $f(s)$ determines the profile of $v_{\rm A}$, it directly affects the propagation and reflection of Alfv\'{e}n waves.
We find that the energy can be transported most efficiently when the distribution of $v_{\rm A}$ along the loop shows a smooth profile.
When a flux tube expands too quickly in altitude, $v_{\rm A}$ gives a dip structure in the low chromosphere. Conversely, when it expands too slowly in altitude, $v_{\rm A}$ shows a bump structure. Concave and bumpy profiles enhance the reflection of Alfv\'{e}n waves so that the coronal heating is suppressed.  As a result, the temperature and density, and accordingly $L_{\rm EUV}$ and $L_{\rm X}$, are lower than those in cases with a smooth  profile of $v_{\rm A}$.
\\

Numerical computations were in part carried out on PC cluster at Center for Computational Astrophysics, National Astronomical Observatory of
Japan.
H.W. is supported by JSPS KAKENHI grant No.JP22J13525.
M.S. is supported by a Grant-in-Aid for Japan Society for the Promotion of Science (JSPS) Fellows and by the NINS program for cross-disciplinary study (grant Nos. 01321802 and 01311904) on Turbulence, Transport, and Heating Dynamics in Laboratory and Solar/Astrophysical Plasmas: “SoLaBo-X.” 
T.K.S. is supported in part by Grants-in-Aid for Scientific Research from the MEXT/JSPS of Japan, 17H01105, 21H00033 and 22H01263 and by Program for Promoting Research on the Supercomputer Fugaku by the RIKEN Center for Computational Science (Toward a unified view of the universe: from large-scale structures to planets; grant 20351188—PI J. Makino) from the MEXT of Japan.

\bibliographystyle{aasjournal} 
\nocite{*}
\bibliography{ref}

\end{document}